\documentclass[conference,compsoc, dvipsnames, table]{IEEEtran}
\usepackage{graphicx}
\usepackage{flushend}
\usepackage{caption}
\usepackage{booktabs}
\usepackage{float}
\usepackage{enumitem}
\usepackage{svg}
\usepackage{xspace}
\usepackage{tabularx}
\usepackage{booktabs}
\usepackage{multirow}
\usepackage{amsmath}
\usepackage{amssymb}
\usepackage{url}

\usepackage{ifpdf}

\ifCLASSOPTIONcompsoc
  \usepackage[nocompress]{cite}
\else
  \usepackage{cite}
\fi
\ifCLASSINFOpdf
\else
\fi
\usepackage{algorithmic}

\usepackage{array}

\ifCLASSOPTIONcompsoc
\begin{document}
%
\title{How Breakable Is Privacy: Probing and Resisting Model Inversion Attacks in Collaborative Inference}

\author{\IEEEauthorblockN{Rongke Liu}
\and
\IEEEauthorblockN{Youwen Zhu}
\and
\IEEEauthorblockN{Dong Wang}
\and
\IEEEauthorblockN{Gaoning Pan}
\and
\IEEEauthorblockN{Xingyu He}
\and
\IEEEauthorblockN{Weizhi Meng}
}
\maketitle

\begin{abstract}
Collaborative inference (CI) improves computational efficiency for edge devices by transmitting intermediate features to cloud models. However, this process inevitably exposes feature representations to model inversion attacks (MIAs), enabling unauthorized data reconstruction. Despite extensive research, there is no established criterion for assessing the difficulty of MIA implementation, leaving a fundamental question unanswered: \textit{What factors truly and verifiably determine the attack's success in CI?} Moreover, existing defenses lack the theoretical foundation described above, making it challenging to regulate feature information effectively while ensuring privacy and minimizing computational overhead. These shortcomings introduce three key challenges: theoretical gap, methodological limitation, and practical constraint.

To overcome these challenges, we propose the first theoretical criterion to assess MIA difficulty in CI, identifying mutual information, entropy, and effective information volume as key influencing factors. The validity of this criterion is demonstrated by using the mutual information neural estimator. Building on this insight, we propose SiftFunnel, a privacy-preserving framework to resist MIA while maintaining usability. Specifically, we incorporate linear and non-linear correlation constraints alongside label smoothing to suppress redundant information transmission, effectively balancing privacy and usability. To enhance deployability, the edge model adopts a funnel-shaped structure with attention mechanisms, strengthening privacy while reducing computational and storage burdens. Experiments show that, compared to state-of-the-art defense, SiftFunnel increases reconstruction error by $\sim$30\%, lowers mutual and effective information metrics by $\geq$50\%, and reduces edge burdens by almost $20\times$, while maintaining comparable usability.
\end{abstract}


%
\IEEEpeerreviewmaketitle

\section{Introduction}
The rapid advancement of deep learning has led to increasingly complex neural networks that achieve remarkable performance in tasks like image recognition and object detection. However, deploying such models on edge devices remains challenging due to limited computational resources~\cite{huckelberry2024tinyml}. Applications like autonomous driving, facial recognition, IoT surveillance, and remote healthcare require real-time processing, yet the limited computational capacity of edge devices makes standalone deep learning inference impractical~\cite{shlezinger2022collaborative}. Cloud computing provides substantial computational power but faces latency, bandwidth, and privacy concerns when transmitting raw data. Collaborative inference (CI)~\cite{shlezinger2021collaborative} mitigates these issues by partitioning neural networks into an edge model \( f_{\text{edge}} \) and a cloud model \( f_{\text{cloud}} \), where \( f_{\text{edge}} \) extracts feature representations for \( f_{\text{cloud}} \) to inference. This method enhances efficiency while maintaining real-time performance and is widely applied in UAVs~\cite{qu2023elastic}, IoT systems~\cite{shlezinger2022collaborative}, and private cloud computing (PCC)~\cite{Apple}. With the rise of 5G, IoT, and AI, CI is poised for broader adoption.

\begin{figure}[t]
    \centering
    \includegraphics[width=1.0\columnwidth]{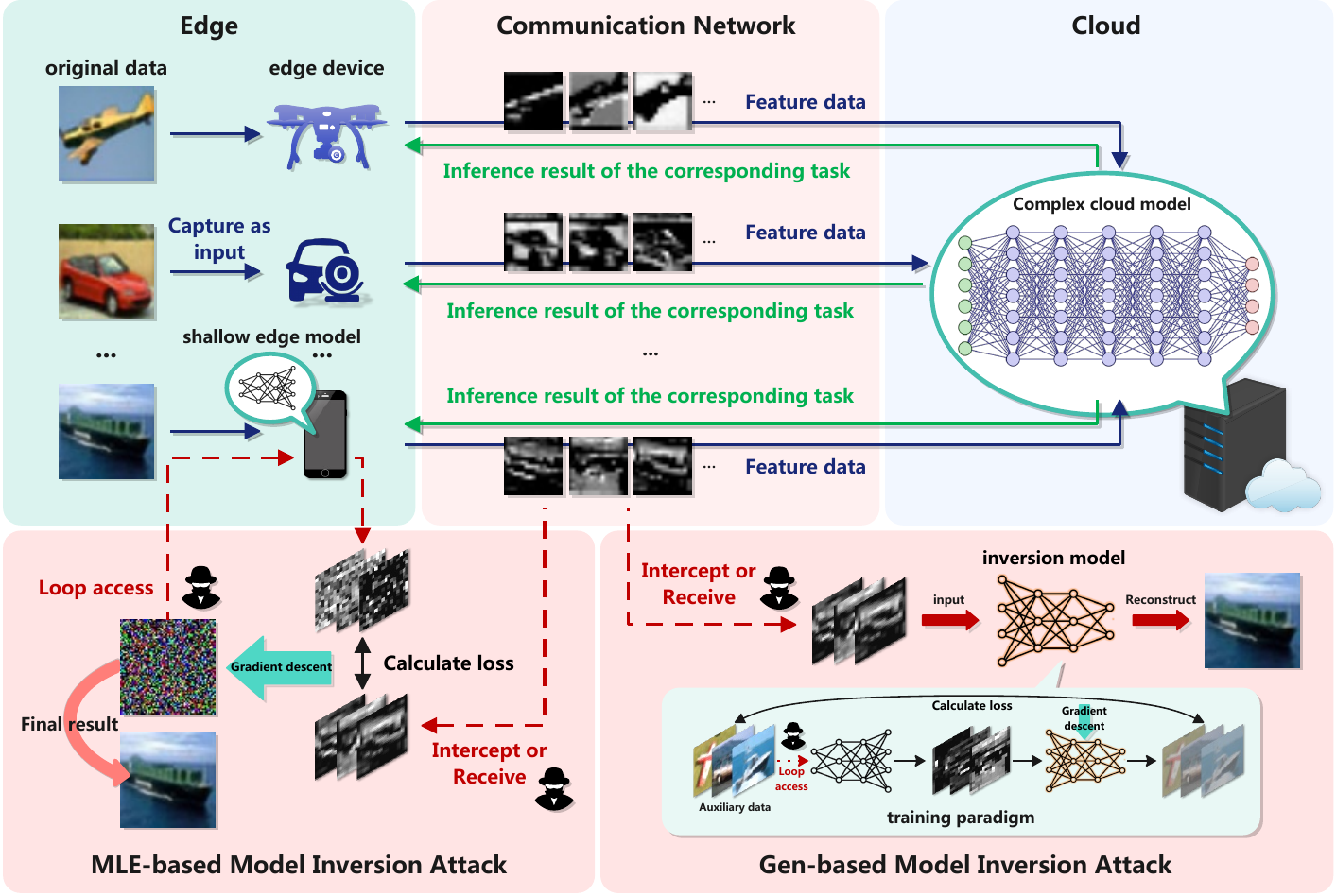}
    \caption{Collaborative inference and the threat of model inversion attacks.}
    \label{fig:collaborative_inference_back}
    \vspace{-0.5em}
\end{figure}

However, recent studies have revealed significant privacy risks in CI, where input data on edge devices can be reconstructed by model inversion attacks (MIAs)~\cite{he2019model, yang2022measuring, yin2023ginver}, as shown in Figure~\ref{fig:collaborative_inference_back}. Unlike centralized deployments, where adversaries recover training data through model APIs~\cite{yuan2023pseudo, struppek2022plug}, CI often relies on shallow neural networks at the edge, producing feature representations that retain significant mutual information (MI) with input data~\cite{azizian2024privacy}. This exposes raw inputs to adversaries, who can reconstruct sensitive data, such as facial images directly from transmitted features~\cite{ding2024patrol}.

MIAs in CI can be categorized into maximum likelihood estimati-on-based (MLE-based) MIAs~\cite{he2019model, he2020attacking} and generative model-based (Gen-based) MIAs~\cite{yin2023ginver, he2020attacking}. MLE-based attacks optimize input data to minimize the loss between the edge model’s output and transmitted features but require white-box access, limiting practicality. Gen-based MIAs, on the other hand, use generative models to approximate the inverse mapping of the edge model, reconstructing inputs from features. Their success depends on data distribution alignment and the model’s ability to learn the inverse function. These privacy risks have driven extensive research on attacks and defenses, but existing methods still face notable limitations.

Firstly, existing research lacks a comprehensive analysis of the fundamental factors influencing MIA effectiveness in CI. Current attack methods focus on improving data reconstruction under specific assumptions, leveraging innovative techniques to enhance reconstruction fidelity and generalization. In recent years, defenses have increasingly recognized the role of mutual information in MIAs and have shifted toward designing optimal information bottleneck (IB) approaches~\cite{wang2024privacy, peng2022bilateral} to counter MIAs. However, these methods remain a methodological extension based on existing findings rather than advancing a fundamental theoretical understanding of MIAs. Consequently, their improvements are inherently constrained by existing technical limitations. Identifying the key factors that control MIA effectiveness and developing a systematic criterion for assessing attack difficulty would significantly deepen our theoretical insights and offer broader theoretical guidance for designing more effective defense strategies.

Secondly, the trade-off between usability, privacy, and deployability in edge models is an underexplored and challenging area in current research. Encryption-based defenses receive little attention in CI due to resource constraints, real-time demands~\cite{mireshghallah2021not}, and vulnerability to side-channel attacks~\cite{hu2023side}. Mainstream MIA defense strategies in CI, constrained by a limited understanding of MIA effectiveness factors, primarily focus on reducing the mutual information \( I(x, z) \) between inputs $x$ and extracted features $z$ to counter MIAs~\cite{wang2024privacy, peng2022bilateral, azizian2024privacy}. However, mutual information estimation is inherently challenging due to its reliance on accurate probability density estimation, large-scale training data, and the computational overhead associated with optimization techniques~\cite{wang2024privacy, duan2023privascissors}, making the usability-privacy trade-off inefficient and difficult to guarantee. Moreover, some approaches compromise usability~\cite{he2020attacking}, increase computational and storage burdens on edge devices~\cite{wang2024privacy}, and complicate further fine-tuning and optimization. Consequently, such methods are often less favored in practical applications.

Thus, there is a pressing need for a novel approach grounded in an in-depth analysis of attack theories to provide interpretable and effective privacy protection while maintaining the usability and deployability of edge models. This need gives rise to three challenges. \textbf{\textit{Challenge 1 (Difficulty assessment):}} \textit{How to establish a verifiable criterion for assessing MIA difficulty in CI?} The fundamental factors determining MIA effectiveness remain poorly understood, making it difficult to develop a reliable assessment criterion. While MI's role in MIAs is recognized, its accurate estimation for validation remains challenging. \textbf{\textit{Challenge 2 (Methodological design):}} \textit{How to effectively and interpretably remove redundant information from transmitted features while preserving edge model usability?} Although IB theory provides a theoretically sound approach, its practical implementation is limited by the technical constraints of MI estimation. Current methods are computationally expensive and lack flexibility, making fine-tuning and further optimization difficult after initial deployment. \textbf{\textit{Challenge 3 (Practical constraint):}} \textit{How to balance usability and privacy while ensuring practical deployability?} Certain defense strategies~\cite{azizian2024privacy} inevitably increase computational and storage burdens on edge devices. Furthermore, when applied to edge models designed for simple feature extraction, these methods struggle to counter advanced Gen-based MIAs effectively.

To overcome the above challenges, we proposed a criterion \( D_{mia} \) to evaluate MIA difficulty in CI and proposed a lightweight privacy-preserving strategy called Information Sifting Funnel (SiftFunnel), featuring 3 key aspects. \textbf{(1) Criterion basis:} \textit{SiftFunnel offers interpretable defense principles based on \( D_{mia} \) .} We analyze MIAs using formulaic principles, information theory, and Fano’s inequality,  and derive an evaluation formula involving 4 critical factors: mutual information \( I(x, z) \), conditional entropy $H(x \mid z)$, feature entropy $H(z)$, and effective information mean \( \delta(z) \) (the average number of non-zero elements in $z$). Their impact on MIA performance is validated via the Mutual Information Neural Estimator (MINE)~\cite{belghazi2018mutual}. \textbf{(2) Usability-privacy trade-off:} \textit{SiftFunnel effectively reduces redundant information while preserving edge model usability.} Aligning with \( D_{mia} \) to strengthen privacy protection, we aim to reduce $H(z)$ and \( \delta(z) \) while increasing $H(x \mid z)$. Specifically, we incorporate distance correlation, Pearson correlation constraints, and an \( l_1 \)-norm constraint on \( z \) to enforce these objectives. To maintain usability, we integrate label smoothing and KL-divergence loss into the training process. This optimization enables the edge model to minify redundant information while ensuring the extraction of task-relevant features. \textbf{(3)  Lightweight design:} \textit{SiftFunnel balances usability and privacy while light-weighting the edge model.} We adopt a funnel-shaped edge model to reduce computational and storage overhead. Attention blocks enhance feature selection, while an upsampling module on the cloud side compensates for lost details, preserving inference accuracy. This design, guided by \( D_{mia} \), also improves privacy by limiting feature \( \delta(z) \) exposure.

Except for previous evaluation metrics, we incorporate MI and \( \delta(z) \) as metrics to assess defense effectiveness. Deployability is measured by the edge model’s parameter count and inference latency. Systematic experiments confirm that SiftFunnel effectively balances usability, deployability, and privacy compared to existing methods, achieving superior MIA resistance while reducing mutual information and storage, with an average accuracy trade-off of 3\%. 

Our contributions are summarized as follows.
\begin{itemize}
    \item We propose the first MIA difficulty criterion in CI, identifying $I(x, z)$, $H(x|z)$, $H(z)$ and $\delta(z)$ as factors impacting attack effectiveness, deepening insights.
    \item We propose SiftFunnel, a privacy-preserving scheme for CI that constrains redundant features via linear and nonlinear correlation, ensuring model usability with label smoothing and loss function regularization.    
    \item We adopt a funnel-shaped structure with attention blocks in SiftFunnel to balance usability and privacy while ensuring lightweight edge model design.
    \item We conduct systematic experiments to validate the proposed criterion and demonstrate SiftFunnel’s effectiveness in balancing usability, privacy, and deployability, using additional metrics including mutual information, $\delta(z)$, and parameter count for evaluation.
\end{itemize}

\section{Background and Related Work}
\subsection{Collaborative inference}

Collaborative inference~\cite{shlezinger2021collaborative} has emerged as an effective solution. This approach divides the inference process into two stages: one part is performed locally on the edge device (which may consist of a single device or a chain of interconnected devices), while the remaining computation is uploaded to the cloud. Formally, this can be expressed as 
\begin{equation}
f(x)=f_e(x) \circ f_c(f_e(x))
\end{equation}
where $f$ is the full trained model, $x$ is the input, $f_e$ represents the model deployed on the edge, and $f_c$ corresponds to the portion deployed in the cloud.

CI typically begins with training $f$ on centralized datasets in the cloud, then partitioning and deploying the $f_e$ to the edge. Therefore, it is imperative to note that CI represents a scenario that has advanced to the inference application phase. Alternatively, approaches like Split Learning ~\cite{zhu2023passive} allow edge devices to participate in model training locally. Regardless of the training approach, the inference process remains consistent. As depicted in Figure~\ref{fig:collaborative_inference_back}, the edge model computes an intermediate representation, which is transmitted to the cloud for completion of the inference process. The final output is either returned to the edge device or retained in the cloud for further analysis.


However, existing research~\cite{he2019model} highlights that edge models are susceptible to MIAs, where adversaries reconstruct input data from transmitted features. Notably, in edge-based inference, computations can be distributed across multiple devices in a chained architecture~\cite{he2019model}, expressed as $f=f_{e1} \circ f_{e2} ... \circ f_{c}$. Studies~\cite{he2020attacking, ho2024model} show that deeper neural networks closer to the cloud transmit more complex, decision-focused features, making them more resistant to reconstruction. In contrast, the initial edge device generates features retaining more input information, making it highly susceptible to MIAs. This paper focuses on the MIA threat to the initial edge device, which employs a shallow neural network while the remaining computation occurs in the cloud. This common scenario in real-world applications presents a significant security risk due to the increased likelihood of input data reconstruction, highlighting the need for effective countermeasures.

\subsection{Model inversion attack}
Current MIAs in CI can be broadly categorized into MLE-based MIAs and Gen-based MIAs, with related research progress described as follows.

\textbf{MLE-based MIAs.} MLE-based MIA was first introduced by Fredrikson et al.~\cite{fredrikson2015model}, who reconstructed training data from a multilayer perceptron (MLP) by minimizing the difference between target and confidence scores using gradient descent. Subsequent research~\cite{he2019model} incorporated Total Variation (TV)~\cite{rudin1992nonlinear} loss as a regularization term to improve optimization, with these advanced techniques referred to as rMLE-based MIA. In this paper, both are collectively referred to as MLE-based MIA. In CI scenarios, MLE-based methods are particularly effective for reconstructing data from shallow neural networks, as the richer feature representations and lower complexity provide adversaries with detailed gradient information that is easier to optimize compared to the prediction vectors of deeper networks.


\textbf{Gen-based MIAs.} This approach was introduced by Yang et al.~\cite{yang2019neural}, who reconstructed input data in facial recognition systems by using generative deep learning models $g$ to approximate the inverse mapping function of the target model $f^{-1}$, such that $g\approx f^{-1}$. By feeding prediction vectors into $g$, they successfully generated facial input images. Building on this approach, He et al.~\cite{he2019model} extended it to collaborative inference, demonstrating that input data from edge models is more vulnerable to reconstruction, thereby highlighting the risks. Unlike MLE-based methods, Gen-based MIA does not require white-box access to the edge model. Instead, it trains an attack model by calculating losses based on transmitted features and auxiliary data, enabling it to approximate the inverse mapping of non-linear computations. Furthermore, as edge models produce rich and redundant outputs, this facilitates the training of attack models and the reconstruction process, as analyzed in detail in Section~\ref{sec:4}.

\textbf{Facts and Motivation.} Although MIA may be confused with “Membership Inference Attack"~\cite{baluta2022membership}, its meaning is clearly defined in this work and widely adopted in the model inversion literature. Moreover, the feasibility of MIA and their threat to CI have been well established. While encryption may protect data in transit, it does not prevent input reconstruction once intermediate features are exposed. Such exposure can occur locally or via compromised cloud infrastructure. Moreover, encryption adds substantial overhead to edge systems, limiting its practicality. These concerns highlight the need to analyze attack factors and design lightweight, deployable defenses.


\subsection{Related defense work}
To safeguard input data collected by edge devices in CI from potential leakage, He et al.~\cite{he2019model, he2020attacking} emphasized the importance of developing effective defense strategies while identifying these vulnerabilities. Existing defenses can be classified into three main categories: perturbation-based, IB-based, and neural network depth (NND) \& IB-based approaches. The following sections outline these strategies.

\textbf{Perturbation-based defense.} This approach introduces carefully designed noise perturbations into the gradients for training or trained edge model output features to reduce the likelihood of adversaries extracting meaningful information and defending against MIA. He et al.~\cite{he2020attacking} initially proposed using Gaussian noise and random feature dropout to limit the effective information accessible to attackers, and later, Wang et al.~\cite{wang2022privacy} extended this strategy by dynamically adjusting the depth of the edge model while adding differential privacy noise to transmitted features. Although these methods enhance defenses, they significantly impact model usability, typically reducing accuracy by approximately 10\%, and remain less effective against Gen-based MIA. The root cause is that the decoder in Gen-based MIAs is sufficiently expressive to adapt to noise or feature dropout, similar to the training of a denoising auto-encoder (DAE)~\cite{gondara2016medical}, potentially increasing attack robustness. In centralized MIA defenses, Struppek et al.~\cite{struppek2023careful} proposed perturbing training labels with a negative label smoothing factor, significantly reducing MIA effectiveness and offering a novel perspective for designing defense strategies.

\textbf{IB-based defense.} This approach enhances defenses against inference attacks by limiting the mutual information between input data $x$ and intermediate features $z$ while preserving the mutual information between $z$ and outputs $y$, thus maintaining inference performance~\cite{wang2021improving}. The objective of this defense can be formally expressed as 
$\min _\theta-{I}(z , y)+ {I}(x, z)$. Initially, Wang et al.~\cite{wang2021improving} introduced information bottleneck theory into MIA defenses, but their method was not optimized for CI and required substantial computational resources for intermediate feature processing. To address this, Peng et al.~\cite{peng2022bilateral} proposed BiDO (Bilateral Dependency Optimization), which uses dependency algorithms to optimize correlations among inputs, features, and outputs. BiDO reduces input-feature correlation while enhancing feature-output correlation, achieving a balance between defense and performance. Additionally, Wang et al.~\cite{wang2024privacy} and Duan et al.~\cite{duan2023privascissors} explored mutual information estimation between $x$ and $z$ using methods based on mutual information definition and the CLUB (Contrastive Log Upper Bound)~\cite{cheng2020club} technique, incorporating the estimates as constraints in the training loss function. IB-based defenses provide interpretable protection mechanisms and valuable design insights for future research. However, these methods face limitations: BiDO relies on kernel functions for non-linear correlation constraints, with performance highly sensitive to hyper-parameters like kernel bandwidth; poor parameter selection can lead to suboptimal results. Methods based on mutual information estimation are constrained by sample scale to compute.

\textbf{NND \& IB-based defense.} Leveraging deeper networks and IB theory, this approach enhances defense effectiveness with targeted loss functions.  Ding et al.~\cite{ding2024patrol} proposed increasing network depth while compressing newly added hidden layers to restrict redundant information transmission. Subsequently, Azizian et al.~\cite{azizian2024privacy} further improved defenses by introducing an auto-encoder (AE) structure, with encoders at the edge and decoders in the cloud. The edge encoder increases network depth and reduces feature dimensionality to achieve the IB effect, while the cloud decoder restores information to maintain predictive performance while applying an $l_{1}$-norm constraint to reduce attack risks. Adversarial training \cite{ding2024patrol} is also incorporated to strengthen robustness against MIAs.

However, this method has certain limitations.  Increasing network depth and adding modules raise computational and hardware demands on edge devices, limiting applicability. Adversarial training, while strengthening defenses, introduces additional challenges—it is less effective against advanced attacks, complicates fine-tuning, and degrades inference performance~\cite{zhang2019limitations}. Additionally, adversaries can adapt their attack models using similar adversarial strategies, weakening defense effectiveness~\cite{jin2024faceobfuscator}. While these methods improve defenses, their implementation requires careful consideration of computational resources, model adaptability, and application suitability. SiftFunnel enhances computational efficiency by deepening edge model structures and adopting a lightweight design, exhibiting IB-like characteristics after training. Thus, our method falls under the NND \& IB-based defense strategy.

\section{Threat Model}
\subsection{Knowledge background of adversary}
\textbf{Attack scenarios. }Based on the adversary's level of knowledge, attack scenarios can be categorized into the following three types:

\begin{enumerate}
    \item \textbf{White-box Scenario:} The adversary has complete white-box knowledge of the edge model, including its architecture, parameters, and gradient information.
    \item \textbf{Black-box Scenario:} The adversary has only black-box access to the model, restricting their capability to observe the intermediate features. Their understanding of the edge model's architecture is also restricted to prior knowledge or incomplete assumptions.
    \item \textbf{Gray-box Scenario:} The adversary has partial knowledge of the edge model's architecture but remains unaware of defensive modifications, such as increased hidden layers.
\end{enumerate}

The gray-box scenario enables the adversary to leverage their partial understanding of the edge model’s architecture to enhance the capability of their attack model. For instance, if the defender partitions the first three convolutional layers of a ResNet-18 model~\cite{he2016deep} and further deepens the network, while the adversary is only aware of the original ResNet-18 architecture, they can only construct an approximate attack model based on partial knowledge of the initial three layers. However, adversaries can address this limitation in this scenario by utilizing their understanding of the edge model’s architecture and reasoning about its mapping capabilities to refine their attack model.

\textbf{Knowledge shared by adversaries in different scenarios. }The adversary is assumed to understand the edge model's task and enhances their attack by training Gen-based MIA on a closely matching data distribution (Section~\ref{sec:6.1}). Additionally, the adversary has access to lossless transmission features and their dimensional distribution, which facilitate optimizing the MIA loss function and training the attack model.

This study focuses on image classification but emphasizes that the defense methods apply to other tasks where the edge model primarily handles feature extraction. Since segmentation, detection, and classification also rely on edge-processed features, MIA and defenses are task-agnostic. Prior work~\cite{azizian2024privacy} has demonstrated input reconstruction and defense effectiveness in object detection using Gen-based MIA.

Moreover, image data is particularly vulnerable to MIAs due to its high dimensionality, as shallow convolutional layers retain more redundant information~\cite{zhang2020secret, he2016deep}. This increases the challenge of protecting image data, making it an ideal test case for privacy defenses. In contrast, textual data, with its strong contextual dependencies and syntactic complexity, has shown limited success in achieving high-quality MIA reconstructions~\cite{dibbo2023sok, zhang2022text}. This study focuses on image data due to its interpretability, making it suitable for evaluating privacy defenses.

\subsection{Attack strategy of the adversary}
\label{sec:3.2}
In the white-box attack scenario, the adversary can perform attacks using the MLE-based MIA method proposed in~\cite{he2019model, he2020attacking}. To ensure the experiments are both comparative and representative, hyper-parameter adjustments were made to achieve performance closer to optimal.

In the black-box attack scenario, the adversary can employ methods from~\cite{yang2019neural, he2019model, yin2023ginver}, relying solely on output features for reverse mapping without access to the edge model's internal details. When protected by NND \& IB-based defenses, the adversary lacks knowledge of architectural changes and must design the attack model based on prior assumptions.

In the gray-box attack scenario, the adversary can employ black-box strategies while leveraging known architectural information to optimize the attack model. This allows the reverse mapping capability to adapt to changes in the edge model. Detailed designs and training procedures for the attack models are provided in Section~\ref{sec:6}.

\section{Theoretical Analysis of MIA in CI}
\label{sec:4}

\subsection{Principles analysis of MIA}
MIAs are categorized into MLE-based and Gen-based methods, both aiming to reconstruct input data using the output of the edge model. MLE-based MIA calculates the loss between transmitted features and the output, then iteratively applies gradient descent to optimize the input, achieving reconstruction of the original data. The attack principle is formulated as follows:
\begin{equation}
\begin{gathered}
x^{(k+1)}=x^{(k)}-\eta \nabla_x \mathcal{L}\left(x, f_{\text {edge }}\left(x^{(k)}\right), z\right), \\
x^*=\arg \min _x \mathcal{L}\left(x, f_{\text {edge }}\left(x^{(k)}\right), z\right)
\end{gathered}
\end{equation}
Where $x^{(k)}$ represents the input value at the $k$-th iteration, typically initialized as Gaussian noise, zeros, or ones. $x^*$ denotes the final reconstructed result, $z$ represents the transmitted features of the target input. $f_{\text {edge }}\left(\cdot\right)$ refers to the target edge model, $\nabla_x \mathcal{L}$ is the gradient calculated from the loss function and edge model parameters to optimize $x$, and $\eta$ is the step size for each update.

According to~\cite{he2019model}, the loss function combines the $l_2$ distance between $f_{\text {edge }}\left(x^{(k)}\right)$ and $z$ with a TV regularization term for $x$, forming the basis of rMLE-based MIA. The method relies on gradient descent, where the loss values depend on output features, and gradients are computed using the target model's parameters.

To achieve effective attacks, sufficient output feature information $\delta(z)$ and a well-chosen loss function are essential to ensure optimization space. Additionally, white-box knowledge of the edge model is required, as complex model mappings and limited feature information increase the risk of gradient descent converging to local minima~\cite{struppek2022plug, zhang2020secret}. For instance, if two similar images produce nearly indistinguishable output features, reconstruction becomes challenging.

When dealing with complex edge model mappings or models protected by defense techniques, directly computing gradients in MLE-based MIA can significantly hinder optimization performance. To address this, He et al.~\cite{he2019model, he2020attacking}, building on the method from~\cite{yang2019neural}, developed a reverse mapping network for the edge model, which is categorized as Gen-based MIA. The key idea is to use a generator to approximate the reverse mapping of the edge model. Different methods adopt various strategies to train the generator. For instance, Yang et al.~\cite{yang2019neural} trained a decoder using auxiliary task-related datasets, while Yin et al.~\cite{yin2023ginver} utilized intercepted feature data for training. The principle of this attack is summarized as follows.
\begin{equation}
\begin{gathered}
G=\arg \min _G \mathbb{E}_{x^{\prime} \sim p\left(x^{\prime}\right)} \mathcal{L}\left(G\left(f_{\text {edge }}\left(x^{\prime}\right)\right), x^{\prime}\right), \\
x^*=G(z)
\end{gathered}
\end{equation}
Where $x^{\prime}$ represents samples from an auxiliary data distribution $p\left(x^{\prime}\right)$. The loss function, typically mean square error, measures the distance between $x^{\prime}$ and the reconstructed data $G\left(f_{\text {edge }}\left(x^{\prime}\right)\right)$ or $z=f_{\text {edge }}\left(x^{\prime}\right)$. The trained generator $G$ must achieve the desired mapping $G \approx {f_{\text{edge}}}^{-1}$ to reconstruct $x^*$.

The above formula highlights the conditions required for effective attacks:
\begin{enumerate}
    \item A well-matched training dataset aligned with the target task, ideally with a distribution similar to the target dataset~\cite{yang2019neural}.
    \item Sufficient and highly separable input information for the generator.
\end{enumerate}

The rationale is that if the information in $f_{\text {edge }}\left(x^{\prime}\right)$ is insufficient or cannot independently represent $x^{\prime}$'s features, the loss in (4) will struggle to converge as $G\left(f_{\text {edge }}\left(x^{\prime}\right)\right)$ remains ambiguous, limiting its training. Among them, the maximum separability is more decisive. For instance, even with random feature drops, if the remaining features distinguish inputs well, the attack can still reconstruct effectively~\cite{he2020attacking}. Such frameworks resemble DAE~\cite{gondara2016medical}, where the decoder can adapt to noise or feature drops.

Gen-based MIA demonstrates superior reconstruction performance compared to MLE-based MIA in defense scenarios. First, it avoids relying on model parameters for gradient computation, eliminating challenges from non-convex optimization. Second, it depends only on the output of the edge model, offering greater flexibility. Furthermore, existing defenses struggle to effectively remove redundant feature information, allowing Gen-based MIA to overcome defenses if the generator is capable of learning the edge model's reverse mapping.

\subsection{Difficulty criterion for MIA in CI}
We summarized the factors influencing MIA effectiveness using a formulaic representation. To further evaluate the difficulty of MIA implementation, we will formalize and analyze these factors from an information-theoretic perspective to establish a clear criterion.

First, the mutual information between the input $x$ and the feature $z$, $I(x; z)$, can be expressed as:
\begin{equation}
I(x ; z)=H(x)-H(x \mid z)=H(z)-H(z \mid x)=H(z)
\end{equation}
Where \(H(x \mid z)\) represents the conditional entropy of input \(x\) given \(z\), and \(H(z \mid x)\) represents the reverse. Since there is no uncertainty in the mapping from \(x\) to \(z\) through \(f_\text{edge}\), \(H(z \mid x)\) is zero. 

This assumption holds primarily because the CI scenario involves deterministic inference, where model parameters are fixed and the feature extractor contains no stochastic components. Even if \(x\) is perturbed by noise, the mapping \(f_{\text{edge}}(x)\) still produces a unique and deterministic output \(z\), ensuring that \(H(z \mid x) = 0\).

\textbf{Factors influencing MLE-based MIA. }From Equation (5), we observe that complex mappings and information loss during neural network forward propagation reduce $I(x; z)$ and increase $H(x|z)$, making non-convex optimization with backpropagation more difficult. Consequently, reconstructing $x$ from $z$ becomes more difficult for MLE-based MIA.

These methods primarily compare $z$ with generated features, but a poorly trained or protected model can produce $z$ with high $\delta(z)$ yet low $H(z)$, indicating suboptimal separability. This weakens attack effectiveness despite the apparent information sufficiency. Thus, MLE-based MIA difficulty is more influenced by $I(x; z)$, $H(x|z)$, and $H(z)$.


To theoretically assess the difficulty of recovering the input $x$, we consider the reconstruction error probability $P_e$. In information theory, Fano’s inequality provides a fundamental lower bound on the reconstruction error in terms of conditional entropy. Modeling the MIA process as a Markov chain $x \to z \to x^*$, where $x^*$ is the reconstruction from $z$, Fano's inequality~\cite{arrow1969classificatory, xu2024query} impiles:
\begin{equation}
H\left(P_e\right)+P_e \log |\mathbb{X}| \geq H(x \mid x^*) \geq H(x \mid z)
\end{equation}
where $\mathbb{X}$ denotes the value space of $x$.


Since $|\mathbb{X}| \geq 2$ and $H(P_e) \leq 1$, we can derive a simplified lower bound:
\begin{equation}
\begin{aligned}
    P_e \geq \frac{H(x \mid z)-1}{\log |\mathbb{X}|}&=\frac{H(x)-I(x,z)-1}{\log |\mathbb{X}|} \\ &=\frac{H(x)-H(z)-1}{\log |\mathbb{X}|}
\end{aligned}
\end{equation}

Here, $H(x)$ and $|\mathbb{X}|$ are considered intrinsic to the dataset and task, and typically remain fixed. Therefore, increasing the lower bound of $P_e$, and hence the difficulty of model inversion, is primarily influenced by $H(x \mid z)$, $H(z)$, and $I(x ; z)$.

While Fano's inequality traditionally assumes exact recovery of $x$, it remains applicable under relaxed semantic criteria, provided that the error probability $P_e$ is redefined accordingly. The inequality merely requires a binary distinction between success and failure, allowing $P_e$ to be interpreted in terms of semantic equivalence or perceptual similarity in the context of model inversion attacks.

\textbf{Factors influencing Gen-based MIA. }From Equation (4), Gen-based MIA requires sufficient input information $\delta(z)$ and high $H(z)$ to ensure maximum separability and enable loss convergence, allowing the generator to simulate reverse mapping effectively. A larger $\delta(z)$ not only facilitates generator training but also directly impacts the reconstruction of $x^*$. At the same time, the relationship in Inequality (7) also holds here~\cite{zhang2023analysis}. Therefore, in addition to the factors influencing MLE-based MIA, Gen-based MIA is further effectively affected by $\delta(z)$.


\textbf{Criterion formula. }The analysis indicates that the difficulty of implementing both types of MIA is inversely proportional to the mutual information, entropy, and output information. Therefore, the difficulty criterion for MIA can be formulated as follows, where $k_1$ and $k_2$ are proportionality constants:
\begin{equation}
D_{\text{mia}} \propto \frac{H(x \mid z)}{H(z)^{k_1} \cdot \delta(z)^{k_2}}
\label{eq:mia-difficulty}, \text{with } I(x; z) = H(z)
\end{equation}

Based on this criterion, existing defenses~\cite{wang2021improving, peng2022bilateral, duan2023privascissors} using the information bottleneck theory can be understood as approximating a reduction in $I(x; z)$ through various algorithms to increase the difficulty of MIA. Perturbation-based defenses~\cite{he2020attacking} aim to directly increase $H(x \mid z)$. The defensive effect of deeper neural networks~\cite{ding2024patrol, azizian2024privacy} can be explained by their ability to reduce $I(x; z)$ and $\delta(z)$. This is mainly because, without skip connections~\cite{hao2025vulnerability}, mutual information progressively decreases during forward propagation in neural networks. In contrast, residual networks with skip connections preserve or even enhance $\delta(z)$ by adding layer outputs to inputs, but the decline in mutual information still poses challenges for MIA (see section~\ref{sec:4.3} for experimental cases).

\subsection{Demonstration and Guidance for Defense}
\label{sec:4.3}
To validate $D_{mia}$ and the theoretical analysis, we designed experiments using a CNN that progressively reduces the spatial size while increasing the channel size, based on the framework in~\cite{yang2019neural, he2016deep}, and ResNet-18 with skip connections, quantifying each block. We trained MINE~\cite{belghazi2018mutual} on the same dataset used to train the full model $f$ to estimate the lower bound of $I(x,z)$. The core idea of MINE is to optimize a neural network estimator using the Donsker-Varadhan (DV)  representation~\cite{donsker1983asymptotic} as the loss function, providing an approximate estimate of the mutual information between random variables. 

Mutual information is defined as follows:
\begin{equation}
I(X ; Z)=\mathbb{E}_{p(x, z)}\left[\log \frac{p(x, z)}{p(x) p(z)}\right]
\end{equation}
Where $p(x, z)$ represents the joint distribution, existing defense methods approximate it, but limited batch sizes lead to inaccuracies and variations in estimates. For high-dimensional data (e.g., images), accurate estimation requires many samples. 

To avoid direct distribution estimation, the DV representation reformulates mutual information as an optimization problem:
\begin{equation}
I(X ; Z) \geq \sup _T \mathbb{E}_{p(x, z)}[T(x, z)]-\log \mathbb{E}_{p(x) p(z)}\left[e^{T(x, z)}\right]
\end{equation}
Where $T$ is a learnable function, this formulation enables $I(X ; Z)$ to be estimated by optimizing $T$ to obtain a lower bound. 

MINE achieves this by parameterizing $T$ with a neural network $T_{\theta}$ and maximizing the DV bound through gradient descent to approximate mutual information:
\begin{equation}
\hat{I}(X ; Z)=\mathbb{E}_{p(x, z)}\left[T_\theta(x, z)\right]-\log \mathbb{E}_{p(x) p(z)}\left[e^{T_\theta(x, z)}\right]
\end{equation}

\begin{figure}[t] 
    \centering
    \includegraphics[width=0.99\columnwidth]{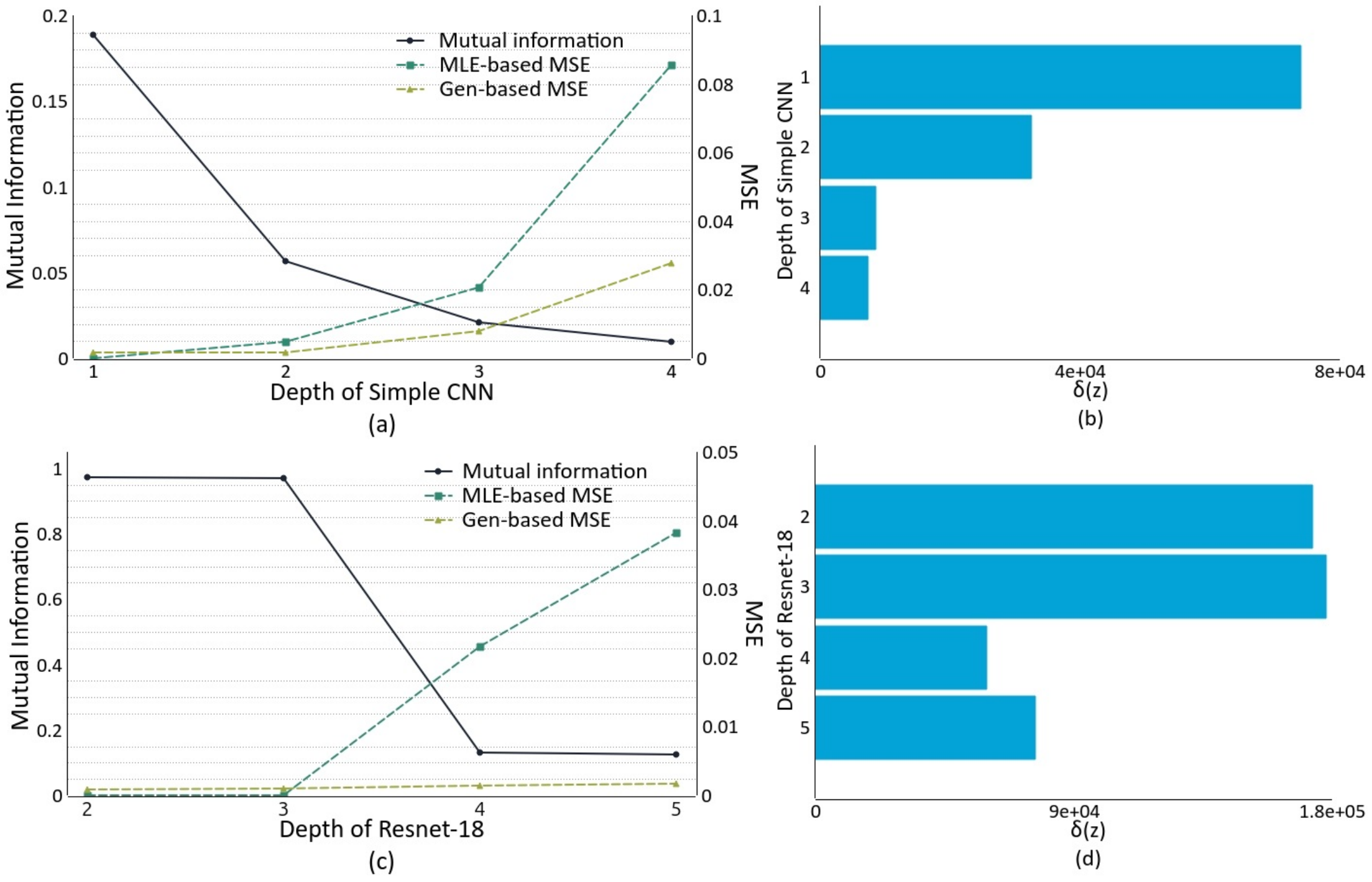}
    \caption{The impact of neural network depth on \textbf{$I(x,z)$, $\delta(z)$}, as well as MLE \& Gen-based MIA.}
    \label{fig:experiment about mi and z}
\end{figure}

As shown in Figure~\ref{fig:experiment about mi and z}, we trained the target models on the CIFAR-10~\cite{krizhevsky2009learning} and used Equation (11) to train MINE for the corresponding blocks. Figures~\ref{fig:experiment about mi and z}(a, b) show that increasing network depth reduces both mutual information and effective information, leading to a decrease in attack performance. Figures~\ref{fig:experiment about mi and z}(c, d) illustrate that while skip connections within two layers do not reduce $\delta(z)$, they slightly lower $I(x,z)$, moderately affecting attack performance. Previous research~\cite{hao2025vulnerability} identified vulnerabilities caused by skip connections in centralized MIA. Our study further highlights their impact on MIA effectiveness in CI.

However, Transmission between blocks in ResNet alters feature channels and dimensions, requiring a shortcut layer, leading to a more pronounced decline in mutual information compared to direct stacking within layers. This, in turn, degrades MIA performance and aligns with our criterion.

\begin{figure*}[t] 
    \centering
    \includegraphics[width=0.84\textwidth]{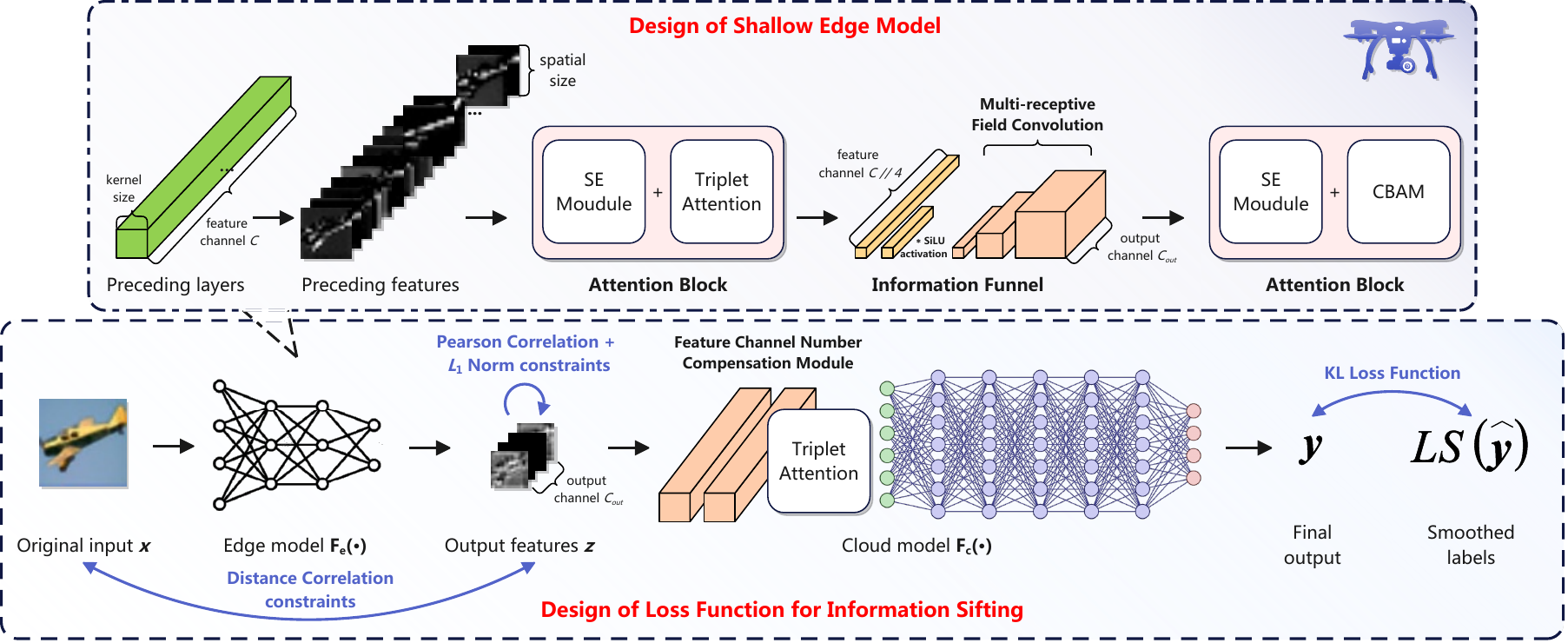}
    \caption{Overview of SiftFunnel. Specifically, the design is divided into two aspects: the architecture and the loss function for the edge and cloud model.}
    \label{fig:SiftFunnel}
\end{figure*}

Notably, while mutual information at ResNet’s fourth layer is higher than that at CNN’s second layer, MLE-based MIA does not perform better. This stems from the fact that these methods rely heavily on gradient descent, making them sensitive to suboptimal optimizers, hyper-parameter settings, and deeper network structures, which hinder gradient calculations for optimizing $x$. The lack of flexible and efficient methods for MLE-based MIA remains a gap in current research, often leading to their exclusion in defense evaluations—a topic warranting further exploration~\cite{ding2024patrol}. In our experiment and Section~\ref{sec:6}, we still evaluate MLE-based MIA against various defenses and provide evidence demonstrating the effectiveness of our method in defending against such attacks.

In conclusion, through theoretical analysis and experimental validation, we proposed $D_{mia}$ and determined that effective defense strategies must impact mutual information, entropy, and $\delta(z)$. The following considerations can guide defense design:
\begin{itemize}
\item[-] Since $I(x,z)$ is challenging to compute and current algorithms lack flexibility and efficiency, defenses should focus on indirectly influencing $H(x\mid z)$ and $H(z)$.
\item[-] The design of edge models should strategically affect $H(z)$ and $\delta(z)$.
\item[-] For practical applications, the light-weighting of edge models should balance usability with their impact on the factors outlined in the criterion. 
\end{itemize}

\section{SiftFunnel: Information Sifting Funnel}

\subsection{Overview of SiftFunnel}
Guided by $D_{{mia}}$, SiftFunnel balances usability and privacy by increasing uncertainty in inferring $x$ from $z$ while limiting redundant information and separability. As shown in Figure~\ref{fig:SiftFunnel}, SiftFunnel introduces two key improvements: First, the edge model modifies the final hidden layer into a funnel-shaped structure and incorporates attention modules. Second, the loss function is optimized to ensure that model training balances usability with privacy protection.

Specifically, in the edge model design, the final layer adopts a funnel-shaped structure, reducing feature channels while preserving spatial dimensions. This reduces computation, lightens the edge device load, and limits redundant information. Attention mechanisms and multi-receptive field convolutions enhance task-relevant information, while a cloud-based upsampling module preserves accuracy. To reduce redundancy, $I(x,z)$ and increase the uncertainty of $z$ to $x$, we use Distance Correlation~\cite{li2012feature} for nonlinear correlations and the Pearson Coefficient~\cite{cohen2009pearson} with $l_1$-norm for linear correlations. Additionally, Label Smoothing (LS) and Kullback-Leibler (KL) divergence~\cite{donsker1983asymptotic} are used during training to improve model usability while reducing the separability of representations in $z$.

\subsection{The Design of Shallow Edge Model}
\label{sec:5.2}
\textbf{Information Funnel. }Deepening the neural network enhances feature extraction and reduces $I(x,z)$, as confirmed in Section~\ref{sec:4}. Previous work~\cite{ho2024model} shows through Fisher information~\cite{rissanen1996fisher} that deeper networks improve task relevance and resist MIA. Inspired by this, we redesign the final layer of the edge model with an Information Funnel module. This module reduces output channels to one-fourth in the first layer and further in the second layer, with channel selection based on the spatial size of feature maps. If the spatial size is reduced significantly, channels can exceed 10 to maintain accuracy. If unchanged, channels can be set to 2 to limit redundancy while preserving accuracy. Adjusting channels based on spatial size is necessary because spatial information loss is irreversible, whereas channel information loss can be compensated in the cloud model. This compensation increases $I(z,y)$ without reversing the reduction in $I(x,z)$, maintaining high attack difficulty as shown in Section~\ref{sec:6.3}. So, we add the channel compensation module in the cloud model to guarantee usability. Importantly, assumptions about extracting features from non-independent transmission modules are idealized, requiring adversaries to have full white-box knowledge of the cloud model as if they were the developer.

The Information Funnel’s channel structure controls output but does not limit redundancy or enhance task-relevant information transmission, affecting usability. Inspired by Inception networks~\cite{szegedy2017inception}, we redesign it as a multi-receptive field convolutional framework with three low-channel layers. A $1*1$ kernel fuses channel information, followed by $3*3$ and $5*5$ kernels to expand the receptive field, extract richer task-relevant features, improve decision-making, and reduce mutual information with $x$, while lowering parameters of the edge model and controlling $\delta(z)$.

\textbf{Attention Block. }To enhance the output of task-relevant information from the Information Funnel, attention modules are incorporated both before and after it. The first block combines an SE block~\cite{hu2018squeeze} and Triplet Attention~\cite{9423300}. The SE block evaluates the contribution of information within each channel, suppressing channels with low task relevance. Triplet Attention reduces spatial information while maintaining smooth, continuous outputs, preserving fine-grained features for further processing. The second block integrates SE and CBAM modules~\cite{woo2018cbam}. Compared to Triplet Attention, CBAM is a lightweight module that sequentially calculates channel and spatial weights to enhance or suppress input features. Since CBAM's channel and spatial attention are independently designed, certain weights are suppressed redundantly, resulting in sparser outputs. This sparsity is beneficial for output processing, effectively increasing $H(x \mid z)$ while reducing $H(z)$ and $\delta(z)$.

\subsection{The Design of Loss Function}
To further enhance model usability and data privacy, appropriately designed loss constraints must be applied to $x$ and $z$ to reasonably increase $H(x \mid z)$ and suppress $H(z)$ within the edge model structure. Without intervention, neural networks may retain irrelevant input information in their output during forward propagation. This information, which does not significantly impact decision-making, reduces the difficulty of MIAs by maintaining the mutual information and entropy. The key challenge is how to achieve this effectively and flexibly.


Some defense approaches utilize Constrained Covariance (COCO) \cite{gretton2005kernel} to estimate correlation, focusing primarily on the linear relationship between $x$ and $z$. However, since neural networks predominantly perform nonlinear operations, studies have shown that this constraint is not only ineffective but may also lead to reduced usability~\cite{peng2022bilateral}. Alternatively, some approaches employ the Hilbert-Schmidt Independency Criterion (HSIC)~\cite{gretton2005measuring} to estimate nonlinear correlation. However, kernel-based calculations in HSIC rely heavily on the appropriate selection of $\sigma$. Improper $\sigma$ value can introduce bias in correlation estimation, limiting the generalizability of this method and reducing its effectiveness across various scenarios.

\textbf{Nonlinear correlation constraint. }We observe that shallow neural network features retain significant similarity to the input and are derived through nonlinear operations. To address this, we use distance correlation to measure the non-linear relationship between $x$ and $z$, incorporating it as a loss constraint. This helps remove redundant input information from the output and effectively increase $H(x\mid z)$. The calculation of distance correlation is as follows:
\begin{equation}
\mathcal{L}_{\text{d Cor}}=\mathrm{d} \operatorname{Cor}(x, z)=\frac{\mathrm{d} \operatorname{Cov}(x, z)}{\sqrt{\mathrm{d} \operatorname{Var}(x) \cdot \mathrm{d} \operatorname{Var}(z)}}
\end{equation}
Where $\mathrm{d} \operatorname{Cov}(x, z)$ represents the distance covariance between $x$ and $z$, and the denominator is the product of the distance variances for the two samples. 

To compute the numerator, a distance matrix $A$ is first calculated, where each element is the Euclidean distance between samples, defined as $A=\left\|x_i-x_j\right\|_2$. To remove the influence of sample shifts, the distance matrix is centered as $A_{i j}^c=A_{i j}-\overline{A_i}-\overline{A_j}+\overline{A_i}$, where $\overline{A_i}$ denotes the mean of the $i$-th column, and similar operations apply to other terms. With the centered distance matrix, the distance covariance is computed as $\mathrm{d} \operatorname{Cov}(x, z)=\frac{1}{\mathrm{n}^2} \sum_{\mathrm{i}, \mathrm{j}} A_{i j}^c B_{i j}^c$, where $B$ represents the matrix of features $z$, and $n$ is the batch size.

The distance variance is defined as $\mathrm{d} \operatorname{Var}(x)=\frac{1}{\mathrm{n}^2} \sum_{\mathrm{i}, \mathrm{j}} A_{i j}^{c^2}$, which normalizes the distance covariance. The value range of distance correlation is $[0,1]$, where $1$ indicates perfect correlation. Since $x$ and $z$ exhibit high distance similarity and are derived through nonlinear operations, distance correlation is well-suited as a constraint term. Moreover, as a distribution-free measure, it does not rely on specific data distributions, making it effective for non-Gaussian distributions, nonlinear relationships, and complex data types such as images and text. For non-stationary and multimodal data, distance correlation provides stable dependency measurements, allowing the constraint to flexibly adapt to different models and data.

\textbf{Linear correlation constraint. } In addition to constraining the correlation between $x$ and $z$, it is necessary to design loss constraints specifically for the $z$. The primary issue lies in the $z$ containing repetitive or similar information, which is redundant for decision-making and provides additional input for Gen-based MIAs, strengthening local constraints in the input space and improving reconstruction accuracy.

To address this, we propose using the Pearson correlation coefficient as a loss constraint to reduce the linear correlation between $x$ and $z$, effectively eliminating redundant information and reducing $\delta(z)$. The calculation is as follows:
\begin{equation}
\begin{aligned}
\mathcal{L}_{\text {Pearson }}&=\frac{1}{C^2} \sum_{c_1=1}^C \sum_{c_2=1}^C\left(\operatorname{p Cor}_{c_1, c_2}\right)^2 \\
&=\frac{1}{C^2} \sum_{c_1=1}^C \sum_{c_2=1}^C\left(\frac{\operatorname{Cov}_{c_1, c_2}}{\sigma_{c_1} \cdot \sigma_{c_2}}\right)^2
\end{aligned}
\end{equation}
Where $\operatorname{Cov}_{c_1, c_2}$ represents the covariance matrix between channels $c_1$ and $c_2$, calculated as $\operatorname{Cov}_{c_1, c_2}=\frac{1}{\mathrm{H} \cdot \mathrm{W}} \sum_{\mathrm{i}=1}^{\mathrm{H} \cdot \mathrm{W}}\left(z_{c_1, i}-\mu_{c_1}\right) \cdot\left(z_{c_2, i}-\mu_{c_2}\right)$, where $z_{c_1, i}$ is the $i$-th element of channel $c_1$ and $\mu_{c_1}$ is its mean. $H$ and $W$ denote the height and width of the feature space. $\sigma_{c_1}$ and $\sigma_{c_2}$ are the standard deviations, calculated as $\sigma_{c_1}=\sqrt{\frac{1}{\mathrm{H} \cdot \mathrm{W}} \sum_{\mathrm{i}=1}^{\mathrm{H} \cdot \mathrm{W}}\left(z_{c_1, i}-\mu_{c_1}\right)^2}$

To compute the scalar $\mathcal{L}_{\text {Pearson }}$, the squared correlations of all channel pairs are averaged, keeping the value range within $[0,1]$. This avoids negative values, which could lead to suboptimal updates during gradient descent. with negative values might drive correlations toward complete negative correlation, introducing unnecessary redundancy or unsuitable learning patterns and hindering convergence. Therefore, the goal is to minimize the coefficient toward zero, as lower values indicate reduced linear correlation. Using the squared correlations and their average ensures a smoother and more stable gradient descent process.

Additionally, to prevent extreme values in the feature output weights during the optimization of the above constraints, we include the $l_1$-norm as a loss term. This term is added specifically to suppress weights, but excessive emphasis could interfere with normal model training. Therefore, the coefficient $\tau$ for this loss term is set below 1e-2 to balance its influence.

\textbf{Label smoothing. }Finally, to ensure model usability, we draw inspiration from~\cite{struppek2023careful} and apply LS to refine the target loss, as shown in formula (14). Additionally, KL divergence is used instead of cross-entropy (CE) as the primary loss function for optimization.
\begin{equation}
L S\left(\hat{y}_c\right)=1-\alpha+\frac{\alpha}{K}, \quad L S\left(\hat{y}_i\right)=\frac{\alpha}{K}
\end{equation}
Where $L S\left(\hat{y}_c\right)$ refers to the label smoothing element under the target class, $L S\left(\hat{y}_i\right)$ is the i-th element of the non-target class. $\alpha$ is smooth factor and $K$ is the total number of categories.

While the $z$ at this layer have not fully transitioned into decision-level semantics, the combination of architectural constraints and the implicit regularization of $I(x,z)$ imposed by the loss function encourages more task-relevant representations. Consequently, positive-factor LS can induce slight intra-class compactness, modestly reducing the feature entropy and hindering inversion attacks~\cite{struppek2023careful}, as evidenced in Section~\ref{sec:6.3}.

Even in cases where LS increases entropy due to insufficient decision-oriented expression, its impact on attack vulnerability remains limited. Ultimately, label smoothing is intended to promote generalization, with improved resistance to MIAs appearing only as a byproduct when sufficient constraints are imposed on the feature representations.


In summary, the proposed loss function first applies nonlinear correlation constraints to increase $H(x| z)$ and suppress $H(z)$. Next, it applies linear correlation and extreme value constraints on $z$ to further enhance $H(x\mid z)$ and suppress $\delta(z)$. Finally, LS and KL divergence maintain usability while slightly suppressing $H(z)$. The final formulation of this optimization is as follows, with specific parameter settings. The final formulation of this optimization is as follows, with specific parameter settings referenced in Section~\ref{sec:6.1}. SiftFunnel and MINE are available at \url{https://github.com/SiftFunnel/SiftFunnel}.
\begin{equation}
\begin{aligned}
    &\min \mathcal{L}_\theta(x, z, \hat{y})=\\
    &\lambda_1 \cdot KL(f(x), LS(\hat{y})) + \lambda_2 \cdot \mathcal{L}_{\text{d Cor}} + \lambda_3 \cdot \mathcal{L}_{\text{Pearson}} + \tau \cdot \|z\|_1
\end{aligned}
\end{equation}

\section{Experiments}
\label{sec:6}
\subsection{Experimental Setup}
\label{sec:6.1}
\textit{1) Datasets. }We used three types of image recognition datasets, with processing details outlined as follows. The detailed data allocation is shown in Table~\ref{tab:data}.

\begin{itemize}
    \item \textbf{CIFAR-10~\cite{krizhevsky2009learning}. }The CIFAR-10 dataset consists of 60,000 RGB images categorized into ten classes with an original resolution of $32 \times 32$. To make the results clearer and more intuitive, we used an unscaled resolution of $64 \times 64$.
    \item \textbf{FaceScrub~\cite{ng2014data}. }FaceScrub is a URL dataset with 100,000 images of 530 actors, which contains 265 male actors and 265 female actors. However, since not every URL was available during the writing period, we downloaded a total of 43,149 images for 530 individuals and resized the images to $64 \times 64$.
    \item \textbf{CelebA~\cite{liu2015deep}. }CelebA is a dataset with 202,599 images of 10,177 celebrities. We used the same crop as~\cite{yang2019neural} to remove the background of images in this dataset other than faces to reduce the impact on 
 the experiment. To eliminate individual overlap, we removed a total of 6,878 images of 296 individuals and similarly resized the images to $64 \times 64$.
    \item \textbf{ChestX-ray~\cite{sait2020curated}. }This dataset is a curated collection of COVID-19 chest X-ray images compiled from 15 publicly available datasets.  It contains 1,281 COVID-19 X-rays, 3,270 normal X-rays, 1,656 viral pneumonia X-rays, and 3,001 bacterial pneumonia X-rays with a resolution of $128 \times 128$.
\end{itemize}

\begin{table}[t]
\caption{Data allocation of the target model and attack model.}
\centering
\resizebox{0.97\columnwidth}{!}{%
\begin{tabular}{@{}cc|c@{}}
\toprule
\multicolumn{2}{c|}{\textbf{Target model}}                                   & \textbf{Attack Model}               \\ \midrule
\multicolumn{1}{c|}{\textbf{Task}}           & \textbf{Data}               & \textbf{Auxiliary Data}             \\ \midrule
\multicolumn{1}{c|}{CIFAR-10 (10 classes)}   & Train: 66.6\%, Test: 16.7\% & 16.7\% (10 classes of CIFAR-10)     \\ \midrule
\multicolumn{1}{c|}{FaceScrub (530 classes)} & Train: 80\%, Test: 20\%     & CelebA (non-individual overlapping) \\ \midrule
\multicolumn{1}{c|}{ChestX-ray (4 classes)}  & Train: 66.6\%, Test: 16.7\% & 16.7\% (4 classes of ChestX-ray)    \\ \bottomrule
\end{tabular}%
}
\label{tab:data}
\end{table}

\begin{table*}[t]
\centering
\caption{Quantitative evaluation of CNN on CIFAR-10 with different defenses against two MIAs. ↑ indicates higher is better, ↓ indicates lower is better. Bolded values are the best. Blue highlights show unprotected baseline performance, red indicates increased edge device load, yellow indicates reduced load, and gray shows gray-box defense performance.}
\resizebox{0.86\textwidth}{!}{
\begin{tabular}{@{}cccccccccccc@{}}
\toprule
                                        &                                                             &                                                                      & \multicolumn{3}{c}{\textbf{MLE-based Attack~\cite{he2019model}}}                                                                                                                                     & \multicolumn{3}{c}{\textbf{Gen-based Attack~\cite{he2020attacking}}}                                                                                &                                                           &                                                          &                                                           \\ \cmidrule(lr){4-9}
\multirow{-2}{*}{\textbf{Method Class}} & \multirow{-2}{*}{\textbf{Method}}                           & \multirow{-2}{*}{\textbf{Test ACC↑}}                                      & \textbf{MSE↑}                                             & \textbf{PSNR↓}                                            & \textbf{SSIM↓}                                            & \textbf{MSE↑}                           & \textbf{PSNR↓}                           & \textbf{SSIM↓}                          & \multirow{-2}{*}{\textbf{MI↓}}                        & \multirow{-2}{*}{\textbf{$\delta(z)$↓}}                         & \multirow{-2}{*}{\textbf{$|\theta_{\text{edge}}|$↓}}                       \\ \midrule
\rowcolor[HTML]{B6DDE8} 
\multicolumn{2}{c}{\cellcolor[HTML]{B6DDE8}Unprotected}                                               & 88.28\%                                                              & 0.0045                                                    & 28.3421                                                   & 0.9883                                                    & 0.0017                                  & 32.5498                                  & 0.9987                                  & 0.0566                                                    & 32,549                                                   & 299,520                                                   \\ \midrule
                                        & \begin{tabular}[c]{@{}c@{}}Adding noise~\cite{he2020attacking} ($\sigma$ = 0.8)\end{tabular}     & 74.85\% \textcolor{Red}{(-13.43\%) }                                                  & 0.0291                                                    & 20.1511                                                   & 0.9135                                                    & 0.0172                                  & 22.4249                                  & 0.9299                                  & 0.0502                                                    & 65,535                                                   &                                                           \\
                                        & \begin{tabular}[c]{@{}c@{}}Dropping~\cite{he2020attacking} (r = 0.8)\end{tabular}   & 21.18\% \textcolor{Red}{(-67.10\%)}                                                   & 0.0127                                                    & 23.7340                                                   & 0.9609                                                    & 0.0135                                  & 23.4835                                  & 0.9481                                  & 0.0501                                                    & 6,509                                                    &                                                           \\
                                        & \begin{tabular}[c]{@{}c@{}}DPSGD~\cite{abadi2016deep} (20, 1e-5)\end{tabular} & 61.83\% \textcolor{Red}{(-26.45\%) }                                                  & 0.0218                                                    & 21.3868                                                   & 0.9479                                                    & 0.0016                                  & 32.4800                                  & 0.994                                   & 0.0623                                                    & 39,829                                                   &                                                           \\
\multirow{-4}{*}{Perturbation-based}    & \begin{tabular}[c]{@{}c@{}}NLS~\cite{struppek2023careful} ($\alpha$ = -0.05)\end{tabular}   & \begin{tabular}[c]{@{}c@{}}85.53\% \textcolor{Red}{(-2.75\%)}\end{tabular}          & 0.0002                                                    & 42.2684                                                   & 0.9996                                                    & 0.0015                                  & 33.0003                                  & 0.9943                                  & 0.0573                                                    & 47,183                                                   & \multirow{-4}{*}{299,520}                                 \\ \midrule
                                        & MID~\cite{wang2021improving} (1e-2)                                                  & 81.53\% \textcolor{Red}{(-6.75\%)}                                                    & 1.74e-5                                                   & 50.4675                                                   & 0.9999                                                    & 0.0013                                  & 34.5184                                  & 0.9967                                  & 0.0584                                                    & 45,198                                                   &                                                           \\
                                        & BiDO~\cite{peng2022bilateral} (2; 20)                                                & 84.12\% \textcolor{Red}{(-4.16\%)}                                                    & 0.6699                                                    & 6.5111                                                    & 0.0138                                                    & 0.035                                   & 19.2965                                  & 0.8751                                  & 0.0429                                                    & 3,345                                                    &                                                           \\
                                        & VIB~\cite{wang2024privacy}                                                         & 88.21\% \textcolor{Red}{(-0.07\%)}                                                    & 0.5552                                                    & 7.3264                                                    & 0.0358                                                    & 0.0042                                  & 28.5054                                  & 0.9842                                  & 0.0551                                                    & 18,069                                                   &                                                           \\
\multirow{-4}{*}{IB-based}               & InfoSCISSORS~\cite{duan2023privascissors}                                                & 83.46\% \textcolor{Red}{(-4.83\%) }                                                   & 0.5715                                                    & 7.2007                                                    & 0.1345                                                    & 0.0062                                  & 26.8799                                  & 0.9948                                  & 0.0549                                                    & 43,364                                                   & \multirow{-4}{*}{299,520}                                 \\ \midrule
                                        & \cellcolor[HTML]{F9DBDF}                                    & \cellcolor[HTML]{F9DBDF}                                             & \cellcolor[HTML]{F9DBDF}                                  & \cellcolor[HTML]{F9DBDF}                                  & \cellcolor[HTML]{F9DBDF}                                  & \cellcolor[HTML]{F9DBDF}0.0542          & \cellcolor[HTML]{F9DBDF}17.4852          & \cellcolor[HTML]{F9DBDF}0.8151          & \cellcolor[HTML]{F9DBDF}                                  & \cellcolor[HTML]{F9DBDF}                                 & \cellcolor[HTML]{F9DBDF}                                  \\
                                        & \multirow{-2}{*}{\cellcolor[HTML]{F9DBDF}AE-based~\cite{azizian2024privacy}}          & \multirow{-2}{*}{\cellcolor[HTML]{F9DBDF}\textbf{90.56\% \textcolor{ForestGreen}{(+2.28\%)}}} & \multirow{-2}{*}{\cellcolor[HTML]{F9DBDF}0.6345}          & \multirow{-2}{*}{\cellcolor[HTML]{F9DBDF}6.7472}          & \multirow{-2}{*}{\cellcolor[HTML]{F9DBDF}0.0051}          & \cellcolor[HTML]{C0C0C0}0.01926         & \cellcolor[HTML]{C0C0C0}21.9259          & \cellcolor[HTML]{C0C0C0}0.9295          & \multirow{-2}{*}{\cellcolor[HTML]{F9DBDF}0.0358}          & \multirow{-2}{*}{\cellcolor[HTML]{F9DBDF}2,617}          & \multirow{-2}{*}{\cellcolor[HTML]{F9DBDF}2,365,056}       \\
                                        & \cellcolor[HTML]{FFFFC7}                                    & \cellcolor[HTML]{FFFFC7}                                             & \cellcolor[HTML]{FFFFC7}                                  & \cellcolor[HTML]{FFFFC7}                                  & \cellcolor[HTML]{FFFFC7}                                  & \cellcolor[HTML]{FFFFC7}\textbf{0.1166} & \cellcolor[HTML]{FFFFC7}\textbf{14.1112} & \cellcolor[HTML]{FFFFC7}\textbf{0.5419} & \cellcolor[HTML]{FFFFC7}                                  & \cellcolor[HTML]{FFFFC7}                                 & \cellcolor[HTML]{FFFFC7}                                  \\
\multirow{-4}{*}{NND \& IB-based}        & \multirow{-2}{*}{\cellcolor[HTML]{FFFFC7}\textbf{SiftFunnel}}     & \multirow{-2}{*}{\cellcolor[HTML]{FFFFC7}85.49\% \textcolor{Red}{(-2.79\%)}}          & \multirow{-2}{*}{\cellcolor[HTML]{FFFFC7}\textbf{0.6792}} & \multirow{-2}{*}{\cellcolor[HTML]{FFFFC7}\textbf{6.4510}} & \multirow{-2}{*}{\cellcolor[HTML]{FFFFC7}\textbf{0.0051}} & \cellcolor[HTML]{C0C0C0}\textbf{0.0639} & \cellcolor[HTML]{C0C0C0}\textbf{16.7314} & \cellcolor[HTML]{C0C0C0}\textbf{0.7647} & \multirow{-2}{*}{\cellcolor[HTML]{FFFFC7}\textbf{0.0167}} & \multirow{-2}{*}{\cellcolor[HTML]{FFFFC7}\textbf{1,006}} & \multirow{-2}{*}{\cellcolor[HTML]{FFFFC7}\textbf{14,911}} \\ \bottomrule
\end{tabular}}
\label{tab:cnn}
\end{table*}

\begin{figure*}[t] 
    \centering
    \includegraphics[width=0.86\textwidth]{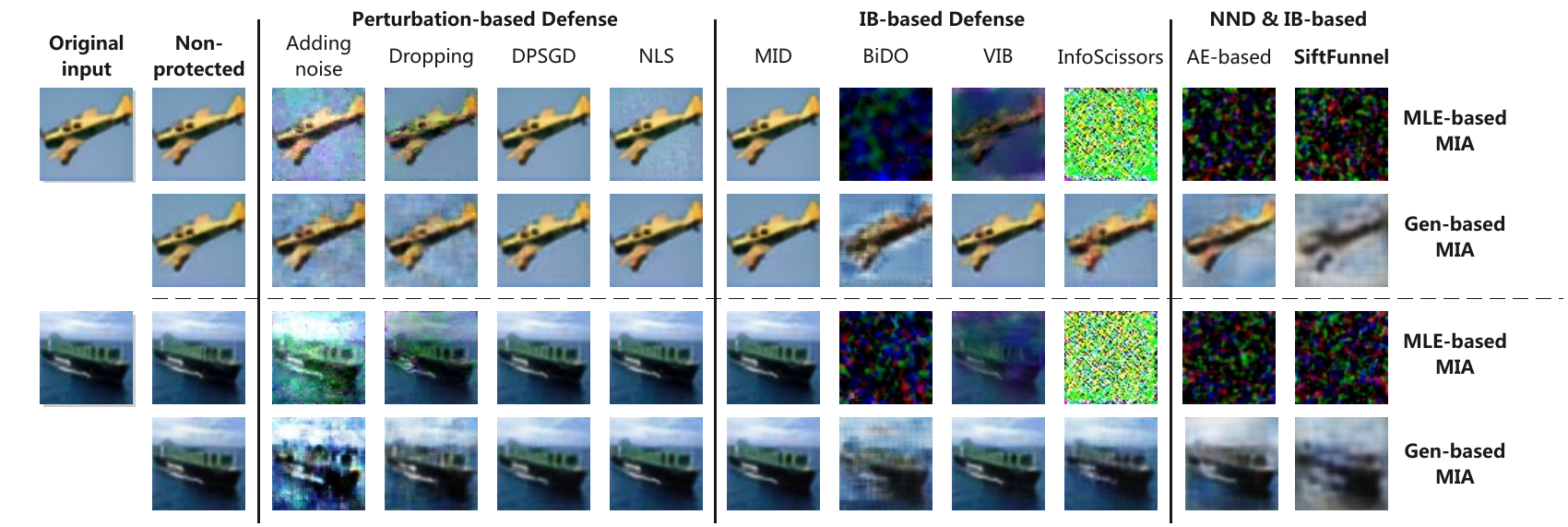}
    \caption{A visual evaluation of the attack effectiveness of two MIAs on CNN edge models protected by various defense techniques, with NND \& IB-based results specifically reflecting the gray-box scenario.}
    \label{fig:er}
\end{figure*}

\textit{2) Attack Method. }The selection of attack methods has been discussed in Section~\ref{sec:3.2}. Specifically, for MLE-based MIA, this paper has adjusted to the hyper-parameter values that yield the best results. For Gen-based MIA, the attack model in this paper is designed as the inverse architecture of the edge model. Moreover, under the gray-box scenario assumption, the attack model can enhance the expressive power of the reverse mapping as the edge model architecture changes. The attack model is trained using the Adam optimizer with a learning rate of 2e-4, a $\beta_1$ of 0.5, and a ReduceLROnPlateaud scheduler with a factor of 0.5 and patience of 20. The attack model training is based on strong assumptions, as shown in Table 1, and only the models with the minimum test MSE are saved to highlight the defense's effectiveness.

\textit{3) Target Edge Model and Implementation Details. }We employ the same comprehensive model architecture as detailed in~\cite{yang2019neural}, comprising four CNN blocks succeeded by two fully connected layers. This model is trained using the Adam optimizer with a learning rate of 2e-4, a $\beta_1$ parameter of 0.5, and a ReduceLROnPlateaud scheduler with a reduction factor of 0.5 and patience of 25. Each CNN block is composed of a convolutional layer, a batch normalization layer, a max-pooling layer, and a ReLU activation function. The target edge model encompasses the initial two CNN blocks. For the training of the edge model protected by SiftFunnel, we have specified the parameters $\lambda_1$ to 3.5, $\alpha$ to 0.35, $\lambda_2$ to 0.8, and $\lambda_3$ to 0.6 in Equation (15). In addition, to argue the impact of the model architecture for the defense, we use different CNNs as follows: (1) VGG16 adapted from~\cite{simonyan2014very}; (2) ResNet-18 adapted from~\cite{he2016deep}, the learning rates are 3e-4. The division positions of ResNet and VGG16 are not fixed. For detailed accuracy and how to divide them, see Section~\ref{sec:6.3} for analysis. All experiments were performed on two RTX 4090 GPUs and an Intel Core i9-14900KF $\times 32$ CPU.

\textit{4) Comparison of Defense Methods. }We aim to compare three existing categories of defense methods. For Perturbation-based defenses, we select \textbf{Adding noise \& Dropping features}~\cite{he2020attacking}, \textbf{DPSGD}~\cite{abadi2016deep}, and negative label smoothing \textbf{(NLS)}~\cite{struppek2023careful}. For IB-based defenses, we choose \textbf{MID}~\cite{wang2021improving}, \textbf{BiDO} (with HSIC)~\cite{peng2022bilateral}, \textbf{VIB}~\cite{wang2024privacy}, and \textbf{InfoSCISSORS}~\cite{duan2023privascissors}. For NND \& IB-based defenses, we opt for \textbf{AE-based protection}~\cite{azizian2024privacy}. Our method falls into the NND \& IB-based defense category, which also requires analyzing the defensive effect in a gray-box scenario. Furthermore, all comparison methods have been selected and set with appropriate hyper-parameters for comparison, with detailed settings found in Section~\ref{sec:6.3}. Lastly, all methods are trained without adversarial training, focusing solely on the effects of the original defense methods.

\vspace{-0.5em}

\subsection{Evaluation Metrics}
We selected 7 evaluation metrics to assess the aspects of usability, privacy, and deployability. For usability assessment, we employ \textbf{Test Accuracy (Test ACC)}, which evaluates the final performance of the model protected by defense techniques on the test dataset. Regarding privacy assessment, we initially utilize \textbf{Mean Squared Error (MSE)}, \textbf{Peak Signal-to-Noise Ratio (PSNR)}, and \textbf{Structural Similarity Index (SSIM)} to evaluate the effectiveness of MIA on reconstructing the input data of the protected model~\cite{yin2023ginver}. Here, a higher MSE and lower PSNR and SSIM (ranging from 0 to 1) indicate better defense effectiveness. In addition, we employ MINE~\cite{belghazi2018mutual}, as mentioned in Section ~\ref{sec:4.3}, to assess \textbf{Mutual Information (MI)} (ranging from 0 to 1) and the \textbf{Effective Information Mean \textbf{$\delta(z)$}}, which is the mean of the non-zero output by the edge model on the test dataset. The specific training methods and model architecture for MINE are referenced from \textit{\url{https://github.com/SiftFunnel/SiftFunnel}}. For deployability, we consider the \textbf{Edge model's parameter count $|\theta_{\text{edge}}|$} and Inference Latency. Model parameters directly affect memory usage, and inference latency is measured on both CPU and GPU.

\begin{table*}[t]
\centering
\caption{Quantitative evaluation of ResNet-18 on CIFAR-10 with different defenses against two MIAs. The target edge model is the first residual block of ResNet-18. ↑ indicates higher is better, ↓ indicates lower is better. Bolded values are the best. The meanings of color-coded regions are consistent with those in Table~\ref{tab:cnn}.}
\resizebox{0.86\textwidth}{!}{
\begin{tabular}{@{}cccccccccccc@{}}
\toprule
                                        &                                                                  &                                                                                                                                            & \multicolumn{3}{c}{\textbf{MLE-based Attack~\cite{he2019model}}}                                                                                                                                     & \multicolumn{3}{c}{\textbf{Gen-based Attack~\cite{he2020attacking}}}                                                                                &                                                           &                                                          &                                                           \\ \cmidrule(lr){4-9}
\multirow{-2}{*}{\textbf{Method Class}} & \multirow{-2}{*}{\textbf{Method}}                                & \multirow{-2}{*}{\textbf{Test ACC↑}}                                                                                                            & \textbf{MSE↑}                                             & \textbf{PSNR↓}                                            & \textbf{SSIM↓}                                            & \textbf{MSE↑}                           & \textbf{PSNR↓}                           & \textbf{SSIM↓}                          & \multirow{-2}{*}{\textbf{MI↓}}                      & \multirow{-2}{*}{\textbf{$\delta(z)$↓}}                  & \multirow{-2}{*}{\textbf{$|\theta_{\text{edge}}|$↓}}      \\ \midrule
\rowcolor[HTML]{B6DDE8} 
\multicolumn{2}{c}{\cellcolor[HTML]{B6DDE8}Unprotected}                                                    & 90.43\%                                                                                                                                    & 0.0001                                                    & 71.0996                                                   & 0.9991                                                    & 0.0010                                  & 34.6599                                  & 0.9991                                  & 0.9719                                                    & 182,134                                                  & 149,824                                                   \\ \midrule
                                        & \begin{tabular}[c]{@{}c@{}}Adding noise~\cite{he2020attacking} ($\sigma$ = 2)\end{tabular}   & \begin{tabular}[c]{@{}c@{}}81.39\% \textcolor{Red}{(-9.04\%)}\end{tabular}                                           & 0.0035                                                    & 29.3354                                                   & 0.9872                                                    & 0.0054                                  & 27.4167                                  & 0.9794                                  & 0.9619                                                    & 262,144                                                  &                                                           \\
                                        & \begin{tabular}[c]{@{}c@{}}Dropping~\cite{he2020attacking} (r = 0.8)\end{tabular}        & \begin{tabular}[c]{@{}c@{}}14.09\% \textcolor{Red}{(-76.34\%)}\end{tabular}                                          & 0.0069                                                    & 26.3641                                                   & 0.9754                                                    & 0.0077                                  & 25.9334                                  & 0.9711                                  & 0.9514                                                    & 35,574                                                   &                                                           \\
                                        & \begin{tabular}[c]{@{}c@{}}DPSGD~\cite{abadi2016deep} (20, 1e-5)\end{tabular}      & \begin{tabular}[c]{@{}c@{}}61.85\% \textcolor{Red}{(-28.58\%)}\end{tabular}                                          & 0.0115                                                    & 23.9453                                                   & 0.9628                                                    & 0.0060                                  & 26.9751                                  & 0.9802                                  & 0.9515                                                    & 178,956                                                  &                                                           \\
\multirow{-4}{*}{Perturbation-based}    & \begin{tabular}[c]{@{}c@{}}NLS~\cite{struppek2023careful} ($\alpha$ = -0.05)\end{tabular} & \begin{tabular}[c]{@{}c@{}}88.13\% \textcolor{Red}{(-2.30\%)}\end{tabular}                                           & 0.0034                                                    & 29.5001                                                   & 0.9894                                                    & 0.0008                                  & 36.0306                                  & 0.9974                                  & 0.9721                                                    & 176,456                                                  & \multirow{-4}{*}{149,824}                                 \\ \midrule
                                        & MID~\cite{wang2021improving} (1e-2)                                                       & \begin{tabular}[c]{@{}c@{}}79.48\% \textcolor{Red}{(-10.95\%)}\end{tabular}                                          & 1.42e-7                                                   & 73.2595                                                   & 0.9999                                                    & 0.0010                                  & 34.5893                                  & 0.9965                                  & 0.9729                                                    & 173,587                                                  &                                                           \\
                                        & BiDO~\cite{peng2022bilateral} (5; 10)                                                     & \textbf{\begin{tabular}[c]{@{}c@{}}90.08\% \textcolor{Red}{(-0.35\%)}\end{tabular}}                                  & 4.38e-7                                                   & 68.3603                                                   & 0.9999                                                    & 0.0020                                  & 31.7621                                  & 0.9927                                  & 0.9647                                                    & 219,832                                                  &                                                           \\
                                        & VIB~\cite{wang2024privacy}                                                              & \begin{tabular}[c]{@{}c@{}}89.18\% \textcolor{Red}{(-1.25\%)}\end{tabular}                                           & 0.5869                                                    & 7.0853                                                    & 0.0216                                                    & 0.0020                                  & 31.7621                                  & 0.9928                                  & 0.9402                                                    & 54,800                                                   &                                                           \\
\multirow{-4}{*}{IB-based}               & InfoSCISSORS~\cite{duan2023privascissors}                                                     & \begin{tabular}[c]{@{}c@{}}88.03\% \textcolor{Red}{(-2.40\%)}\end{tabular}                                           & 3.10e-5                                                   & 49.8548                                                   & 0.9999                                                    & 0.0058                                  & 27.1608                                  & 0.9799                                  & 0.8927                                                    & 216,919                                                  & \multirow{-4}{*}{149,824}                                 \\ \midrule
                                        & \cellcolor[HTML]{F9DBDF}                                         & \cellcolor[HTML]{F9DBDF}                                                                                                                   & \cellcolor[HTML]{F9DBDF}                                  & \cellcolor[HTML]{F9DBDF}                                  & \cellcolor[HTML]{F9DBDF}                                  & \cellcolor[HTML]{F9DBDF}0.0143          & \cellcolor[HTML]{F9DBDF}23.2332          & \cellcolor[HTML]{F9DBDF}0.9488          & \cellcolor[HTML]{F9DBDF}                                  & \cellcolor[HTML]{F9DBDF}                                 & \cellcolor[HTML]{F9DBDF}                                  \\
                                        & \multirow{-2}{*}{\cellcolor[HTML]{F9DBDF}AE-based~\cite{azizian2024privacy}}               & \multirow{-2}{*}{\cellcolor[HTML]{F9DBDF}\begin{tabular}[c]{@{}c@{}}89.19\% \textcolor{Red}{(-1.24\%)}\end{tabular}} & \multirow{-2}{*}{\cellcolor[HTML]{F9DBDF}0.6543}          & \multirow{-2}{*}{\cellcolor[HTML]{F9DBDF}6.6189}          & \multirow{-2}{*}{\cellcolor[HTML]{F9DBDF}0.1243}          & \cellcolor[HTML]{C0C0C0}0.0094          & \cellcolor[HTML]{C0C0C0}25.0633          & \cellcolor[HTML]{C0C0C0}0.9660          & \multirow{-2}{*}{\cellcolor[HTML]{F9DBDF}0.9397}          & \multirow{-2}{*}{\cellcolor[HTML]{F9DBDF}74,356}         & \multirow{-2}{*}{\cellcolor[HTML]{F9DBDF}279,136}         \\
                                        & \cellcolor[HTML]{FFFFC7}                                         & \cellcolor[HTML]{FFFFC7}                                                                                                                   & \cellcolor[HTML]{FFFFC7}                                  & \cellcolor[HTML]{FFFFC7}                                  & \cellcolor[HTML]{FFFFC7}                                  & \cellcolor[HTML]{FFFFC7}\textbf{0.1957} & \cellcolor[HTML]{FFFFC7}\textbf{11.8562} & \cellcolor[HTML]{FFFFC7}\textbf{0.0793} & \cellcolor[HTML]{FFFFC7}                                  & \cellcolor[HTML]{FFFFC7}                                 & \cellcolor[HTML]{FFFFC7}                                  \\
\multirow{-4}{*}{NND \& IB-based}        & \multirow{-2}{*}{\cellcolor[HTML]{FFFFC7}\textbf{SiftFunnel}}          & \multirow{-2}{*}{\cellcolor[HTML]{FFFFC7}88.43\% \textcolor{Red}{(-2.00\%)}}                                            & \multirow{-2}{*}{\cellcolor[HTML]{FFFFC7}\textbf{0.6538}} & \multirow{-2}{*}{\cellcolor[HTML]{FFFFC7}\textbf{6.6166}} & \multirow{-2}{*}{\cellcolor[HTML]{FFFFC7}\textbf{0.0059}} & \cellcolor[HTML]{C0C0C0}\textbf{0.2790} & \cellcolor[HTML]{C0C0C0}\textbf{10.3230} & \cellcolor[HTML]{C0C0C0}\textbf{0.3001} & \multirow{-2}{*}{\cellcolor[HTML]{FFFFC7}\textbf{0.4772}} & \multirow{-2}{*}{\cellcolor[HTML]{FFFFC7}\textbf{4,028}} & \multirow{-2}{*}{\cellcolor[HTML]{FFFFC7}\textbf{78,864}} \\ \bottomrule
\end{tabular}}
\label{tab:res}
\end{table*}

\begin{table*}[t]
\caption{Quantitative evaluation of different defense methods in protecting CNN against MIAs on the FaceScrub dataset. Changes in data types alter the inference task, requiring adjustments to the parameters of the defense methods.}
\centering
\resizebox{0.86\textwidth}{!}{%
\begin{tabular}{@{}cccccccccccc@{}}
\toprule
 &
   &
   &
  \multicolumn{3}{c}{\textbf{MLE-based Attack~\cite{he2019model}}} &
  \multicolumn{3}{c}{\textbf{Gen-based Attack~\cite{he2020attacking}}} &
   &
   &
   \\ \cmidrule(lr){4-9}
\multirow{-2}{*}{\textbf{Method Class}} &
  \multirow{-2}{*}{\textbf{Method}} &
  \multirow{-2}{*}{\textbf{Test ACC↑}} &
  \textbf{MSE↑} &
  \textbf{PSNR↓} &
  \textbf{SSIM↓} &
  \textbf{MSE↑} &
  \textbf{PSNR↓} &
  \textbf{SSIM↓} &
  \multirow{-2}{*}{\textbf{MI↓}} &
  \multirow{-2}{*}{\textbf{$\delta(z)$↓}} &
  \multirow{-2}{*}{\textbf{$|\theta_{\text{edge}}|$↓}} \\ \midrule
\rowcolor[HTML]{B6DDE8} 
\multicolumn{2}{c}{\cellcolor[HTML]{B6DDE8}Unprotected} &
  88.31\% &
  0.0005 &
  37.6115 &
  0.9972 &
  0.0006 &
  32.4108 &
  0.9908 &
  0.0514 &
  45,271 &
  299,520 \\ \midrule
Perturbation-based &
  \begin{tabular}[c]{@{}c@{}}NLS~\cite{struppek2023careful} ($\alpha$ = -0.2)\end{tabular} &
  88.03\% \textcolor{Red}{(-0.28\%)} &
  0.0004 &
  38.8904 &
  0.9980 &
  0.0017 &
  32.5480 &
  0.9911 &
  0.0515 &
  44,783 &
  299,520 \\ \midrule
 &
MID~\cite{wang2021improving} (1e-2) &
  79.45\% \textcolor{Red}{(-10.98\%)} &
  0.0017 &
  32.3435 &
  0.9897 &
  0.0018 &
  32.1285 &
  0.9901 &
  0.0507 &
  39,277 &
   \\
 &
  BiDO~\cite{peng2022bilateral} (2; 20) &
  88.02\% \textcolor{Red}{(-0.29\%)} &
  0.0004 &
  39.0838 &
  0.9978 &
  0.0018 &
  32.2048 &
  0.9904 &
  0.0493 &
  44,915 &
   \\
 &
  VIB~\cite{wang2024privacy} &
  \textbf{88.05\% \textcolor{Red}{(-0.26\%)}} &
  0.7354 &
  6.1060 &
  0.0032 &
  0.0033 &
  29.5697 &
  0.9823 &
  0.0493 &
  15,649 &
   \\
\multirow{-4}{*}{IB-based} &
  InfoSCISSORS~\cite{duan2023privascissors} &
  81.61\% \textcolor{Red}{(-6.70\%)} &
  \textbf{0.7449} &
  \textbf{6.0504} &
  \textbf{0.0026} &
  0.0060 &
  26.9803 &
  0.9675 &
  0.0408 &
  57,050 &
  \multirow{-4}{*}{299,520} \\ \midrule
 &
  \cellcolor[HTML]{F9DBDF} &
  \cellcolor[HTML]{F9DBDF} &
  \cellcolor[HTML]{F9DBDF} &
  \cellcolor[HTML]{F9DBDF} &
  \cellcolor[HTML]{F9DBDF} &
  \cellcolor[HTML]{F9DBDF}\textbf{0.0609} &
  \cellcolor[HTML]{F9DBDF}\textbf{16.9260} &
  \cellcolor[HTML]{F9DBDF}\textbf{0.6656} &
  \cellcolor[HTML]{F9DBDF} &
  \cellcolor[HTML]{F9DBDF} &
  \cellcolor[HTML]{F9DBDF} \\
 &
  \multirow{-2}{*}{\cellcolor[HTML]{F9DBDF}AE-based~\cite{azizian2024privacy}} &
  \multirow{-2}{*}{\cellcolor[HTML]{F9DBDF}83.45 \textcolor{Red}{(-4.86\%)}} &
  \multirow{-2}{*}{\cellcolor[HTML]{F9DBDF}0.6474} &
  \multirow{-2}{*}{\cellcolor[HTML]{F9DBDF}6.6594} &
  \multirow{-2}{*}{\cellcolor[HTML]{F9DBDF}0.0090} &
  \cellcolor[HTML]{C0C0C0}0.0220 &
  \cellcolor[HTML]{C0C0C0}21.3444 &
  \cellcolor[HTML]{C0C0C0}0.8795 &
  \multirow{-2}{*}{\cellcolor[HTML]{F9DBDF}0.0238} &
  \multirow{-2}{*}{\cellcolor[HTML]{F9DBDF}\textbf{585}} &
  \multirow{-2}{*}{\cellcolor[HTML]{F9DBDF}2,365,056} \\
 &
  \cellcolor[HTML]{FFFFC7} &
  \cellcolor[HTML]{FFFFC7} &
  \cellcolor[HTML]{FFFFC7} &
  \cellcolor[HTML]{FFFFC7} &
  \cellcolor[HTML]{FFFFC7} &
  \cellcolor[HTML]{FFFFC7}0.0559 &
  \cellcolor[HTML]{FFFFC7}17.3018 &
  \cellcolor[HTML]{FFFFC7}0.6854 &
  \cellcolor[HTML]{FFFFC7} &
  \cellcolor[HTML]{FFFFC7} &
  \cellcolor[HTML]{FFFFC7} \\
\multirow{-4}{*}{NND \& IB-based} &
  \multirow{-2}{*}{\cellcolor[HTML]{FFFFC7}\textbf{SiftFunnel ($\lambda_1=5.0,\lambda_2=0.1,\lambda_3=0.1$)}} &
  \multirow{-2}{*}{\cellcolor[HTML]{FFFFC7}85.39\% \textcolor{Red}{(-2.92\%)}} &
  \multirow{-2}{*}{\cellcolor[HTML]{FFFFC7}0.7264} &
  \multirow{-2}{*}{\cellcolor[HTML]{FFFFC7}6.1593} &
  \multirow{-2}{*}{\cellcolor[HTML]{FFFFC7}0.0060} &
  \cellcolor[HTML]{C0C0C0}\textbf{0.0296} &
  \cellcolor[HTML]{C0C0C0}\textbf{20.0576} &
  \cellcolor[HTML]{C0C0C0}\textbf{0.8323} &
  \multirow{-2}{*}{\cellcolor[HTML]{FFFFC7}\textbf{0.0104}} &
  \multirow{-2}{*}{\cellcolor[HTML]{FFFFC7}637} &
  \multirow{-2}{*}{\cellcolor[HTML]{FFFFC7}\textbf{14,911}} \\ \bottomrule
\end{tabular}}
\label{tab:face}
\end{table*}

\subsection{Experimental Results}
\label{sec:6.3}

\vspace{-0.5em}
The results are analyzed from 4 perspectives: the impact of defense methods on $I(x,z)$ and $\delta(z)$, the influence of datasets and model architectures, and ablation studies.

\textit{1) The effect of various defenses on the MI and $\delta(z)$. }As shown in Table~\ref{tab:cnn}, defense effectiveness is reflected by MI and $\delta(z)$, aligning with the criterion $D_{\text{mia}}$. Specifically, unprotected edge models are highly vulnerable to MIAs, with MI at 0.0566, MSE below 0.01, and SSIM exceeding 0.98. Perturbation-based defenses reduce MI by at most 11\% but cause over a 10\% drop in Test ACC, with NLS even increasing MI in shallow networks, weakening resistance to attacks. From the theoretical analysis, it can be seen that DPSGD does not alter the complexity of the mapping from $x$ to $z$, nor does it directly affect the intermediate feature $z$ itself. As a result, it reduces model accuracy without effectively mitigating MIA risks in the CI setting. IB-based defenses slightly reduce usability but effectively counter MLE-based MIAs, except for MID, which shows weaker resistance. Among them, BiDO significantly impacts MI and $\delta(z)$, providing effective defense, but its parameter selection and kernel width configuration (as shown in the table) limits generalizability, as confirmed in subsequent experiments. This experiment follows the InfoSCISSORS setup, ensuring CLUB receives a complete dataset and sufficient training for convergence. While CLUB effectively defends against MLE-based MIAs, it struggles against Gen-based MIAs and incurs a 4.83\% accuracy loss. AE-based defenses improve model usability, due to enhanced model capacity, while MI reductions confirm reasonable defense effectiveness. However, gray-box scenarios weaken AE-based defenses, and their added complexity increases storage costs by $8\times$, as seen in $|\theta_{\text{edge}}|$. SiftFunnel achieves the best defense performance against both MIAs while keeping accuracy loss below 3\%. It achieves MSE of 0.6792 (MLE-based MIA) and above 0.1 (Gen-based MIA). In gray-box scenarios, it maintains MSE of 0.06 and SSIM of 0.7647, ensuring robust resistance. Furthermore, MI and $\delta(z)$ confirm its optimal impact on the key factors of $D_{\text{mia}}$. Additionally, $|\theta_{\text{edge}}|$ reveals a nearly $20\times$ reduction in edge model load, highlighting SiftFunnel’s efficiency.

\begin{table*}[t]
\centering
\caption{SiftFunnel Ablation Study on CIFAR-10. The target edge model is the first residual block of ResNet-18, with Gen-based MIA as the attack method. The Red area indicates the strong attack hypothesis from Section~\ref{sec:5.2}.}
\resizebox{0.81\textwidth}{!}{
\begin{tabular}{@{}ccccccc@{}}
\toprule
\textbf{Method} &
  \textbf{Test ACC↑} &
  \textbf{MSE↑} &
  \textbf{PSNR↓} &
  \textbf{SSIM↓} &
  \textbf{MI↓} &
  \textbf{$\delta$(z)↓} \\ \midrule
\rowcolor[HTML]{B6DDE8} 
Unprotected &
  90.43\% &
  0.0010 &
  34.6599 &
  0.9991 &
  0.9719 &
  182,134 \\ \midrule
Without Funnel (Average training round takes 1 minute, batchsize=64) &
  \textbf{90.33\% \textcolor{Red}{(-0.10\%)}} &
  0.0594 &
  17.0709 &
  0.7776 &
  0.8945 &
  24,322 \\ \midrule
Without Attention Blocks &
  88.45\% \textcolor{Red}{(-1.98\%)} &
  0.0525 &
  17.5709 &
  0.8178 &
  0.8763 &
  \textbf{3,904} \\ \midrule
Without $KL$ and $LS$ &
  88.59\% \textcolor{Red}{(-1.84\%)} &
  0.0850 &
  15.4813 &
  0.6905 &
  0.7935 &
  3,994 \\ \midrule
Without $\mathcal{L}_{\text{dCor}}$ &
  88.77\% \textcolor{Red}{(-1.66\%)} &
  0.0268 &
  20.4929 &
  0.9025 &
  0.9505 &
  4,013 \\ \midrule
Without $\mathcal{L}_{\text{Pearson}}$ &
  88.83\% \textcolor{Red}{(-1.60\%)} &
  0.0492 &
  17.8576 &
  0.8294 &
  0.8618 &
  4,037 \\ \midrule
Without $l_1$ &
  88.48\% \textcolor{Red}{(-1.95\%)} &
  0.0789 &
  15.8057 &
  0.7272 &
  0.8158 &
  4,030 \\ \midrule
Attack SiftFunnel in Cloud &
   &
  \cellcolor[HTML]{F9DBDF}0.0521 &
  \cellcolor[HTML]{F9DBDF}17.6159 &
  \cellcolor[HTML]{F9DBDF}0.8205 &
  \cellcolor[HTML]{F9DBDF}0.8905 &
  \cellcolor[HTML]{F9DBDF}176,380 \\ \cmidrule(r){1-1} \cmidrule(l){3-7} 
SiftFunnel &
  \multirow{-2.6}{*}{88.43\% \textcolor{Red}{(-2.00\%)}} &
  \cellcolor[HTML]{FFFFFF}\textbf{0.2790} &
  \cellcolor[HTML]{FFFFFF}\textbf{10.3230} &
  \cellcolor[HTML]{FFFFFF}\textbf{0.3001} &
  \cellcolor[HTML]{FFFFFF}\textbf{0.4772} &
  \cellcolor[HTML]{FFFFFF}4,028 \\ \bottomrule
\end{tabular}}
\label{tab:ablation}
\end{table*}

\textit{2) The effect of model architecture. }Modifying the model architecture does not alter the parameter settings of the proposed method, which remain consistent with Section~\ref{sec:6.1}. We adapt the model to ResNet-18 with skip connections, using the first residual block as the edge model. This setup poses challenges for defense techniques, as shallow networks retain simple structures and do not modify spatial dimensions. Table~\ref{tab:res} shows that unprotected models exhibit minimal MI impact, resulting in weak resistance to MIAs, with MLE-based MIA achieving an MSE of $10^{-4}$ and Gen-based MIA $10^{-3}$. Under these conditions, perturbation-based defenses fail to balance usability and privacy, leading to poor defense performance. Among IB-based defenses, only VIB demonstrates resistance to MLE-based MIA by optimizing the edge model through $I(x,z)$ estimation as a loss function, but its effectiveness remains limited, particularly against Gen-based MIA. For BiDO, despite applying kernel bandwidth estimation methods from~\cite{peng2022bilateral} and extensive experimental tuning, it fails to achieve a configuration that effectively balances usability and privacy, resulting in suboptimal defense performance. InfoSCISSORS impacts MI under this configuration, but the increase in $\delta(z)$ undermines its effectiveness against Gen-based MIAs, leading to only marginal improvements. AE-based defenses show some resistance to attacks on low-dimensional features in earlier stages, but their effectiveness diminishes under the ResNet configuration. While they reduce $\delta(z)$ by 60\%, they have minimal impact on MI, achieving only a 0.04\% reduction. Consequently, AE-based defenses struggle to counter both black-box and gray-box Gen-based MIAs effectively.

\begin{table}[t]
\centering
\caption{Quantitative evaluation of MIA resistance on CIFAR-10 across different architectures. Edge models for CNN and ResNet-18 follow Tables~\ref{tab:cnn} and~\ref{tab:res}; for VGG16, the split is placed at the first ReLU after max pooling.}
\resizebox{0.99\columnwidth}{!}{
\begin{tabular}{@{}ccccccccccc@{}}
\toprule
 &
   &
  \multicolumn{3}{c}{\textbf{MLE-based Attack~\cite{he2019model}}} &
  \multicolumn{3}{c}{\textbf{Gen-based Attack~\cite{he2020attacking}}} &
   &
   &
   \\ \cmidrule(lr){3-8}
\multirow{-2}{*}{\textbf{Architecture}} &
  \multirow{-2}{*}{\textbf{Test ACC↑}} &
  \textbf{MSE↑} &
  \textbf{PSNR↓} &
  \textbf{SSIM↓} &
  \textbf{MSE↑} &
  \textbf{PSNR↓} &
  \textbf{SSIM↓} &
  \multirow{-2}{*}{\textbf{MI↓}} &
  \multirow{-2}{*}{\textbf{$\delta(z)$↓}} &
  \multirow{-2}{*}{\textbf{$|\theta_{\text{edge}}|$↓}} \\ \midrule
\rowcolor[HTML]{FFFFFF} 
\cellcolor[HTML]{FFFFFF} &
  \cellcolor[HTML]{FFFFFF} &
  \cellcolor[HTML]{FFFFFF} &
  \cellcolor[HTML]{FFFFFF} &
  \cellcolor[HTML]{FFFFFF} &
  0.1166 &
  14.1112 &
  0.5419 &
  \cellcolor[HTML]{FFFFFF} &
  \cellcolor[HTML]{FFFFFF} &
  \cellcolor[HTML]{FFFFFF} \\
\rowcolor[HTML]{FFFFFF} 
\multirow{-2}{*}{\cellcolor[HTML]{FFFFFF}CNN} &
  \multirow{-2}{*}{\cellcolor[HTML]{FFFFFF}85.49\% \textcolor{Red}{(-2.79\%)}} &
  \multirow{-2}{*}{\cellcolor[HTML]{FFFFFF}0.6792} &
  \multirow{-2}{*}{\cellcolor[HTML]{FFFFFF}6.4510} &
  \multirow{-2}{*}{\cellcolor[HTML]{FFFFFF}0.0051} &
  \cellcolor[HTML]{C0C0C0}0.0639 &
  \cellcolor[HTML]{C0C0C0}16.7314 &
  \cellcolor[HTML]{C0C0C0}0.7647 &
  \multirow{-2}{*}{\cellcolor[HTML]{FFFFFF}0.0167} &
  \multirow{-2}{*}{\cellcolor[HTML]{FFFFFF}1,006} &
  \multirow{-2}{*}{\cellcolor[HTML]{FFFFFF}14,911} \\
 &
  \cellcolor[HTML]{FFFFFF} &
  \cellcolor[HTML]{FFFFFF} &
  \cellcolor[HTML]{FFFFFF} &
  \cellcolor[HTML]{FFFFFF} &
  \cellcolor[HTML]{FFFFFF}0.1957 &
  \cellcolor[HTML]{FFFFFF}11.8562 &
  \cellcolor[HTML]{FFFFFF}0.0793 &
  \cellcolor[HTML]{FFFFFF} &
  \cellcolor[HTML]{FFFFFF} &
  \cellcolor[HTML]{FFFFFF} \\
\multirow{-2}{*}{ResNet-18} &
  \multirow{-2}{*}{\cellcolor[HTML]{FFFFFF}88.43\% \textcolor{Red}{(-2.00\%)}} &
  \multirow{-2}{*}{\cellcolor[HTML]{FFFFFF}0.6538} &
  \multirow{-2}{*}{\cellcolor[HTML]{FFFFFF}6.6166} &
  \multirow{-2}{*}{\cellcolor[HTML]{FFFFFF}0.0059} &
  \cellcolor[HTML]{C0C0C0}0.2790 &
  \cellcolor[HTML]{C0C0C0}10.3230 &
  \cellcolor[HTML]{C0C0C0}0.3001 &
  \multirow{-2}{*}{\cellcolor[HTML]{FFFFFF}0.4772} &
  \multirow{-2}{*}{\cellcolor[HTML]{FFFFFF}4,028} &
  \multirow{-2}{*}{\cellcolor[HTML]{FFFFFF}78,864} \\
 &
  \cellcolor[HTML]{FFFFFF} &
  \cellcolor[HTML]{FFFFFF} &
  \cellcolor[HTML]{FFFFFF} &
  \cellcolor[HTML]{FFFFFF} &
  \cellcolor[HTML]{FFFFFF}0.0827 &
  \cellcolor[HTML]{FFFFFF}15.5928 &
  \cellcolor[HTML]{FFFFFF}0.6559 &
  \cellcolor[HTML]{FFFFFF} &
  \cellcolor[HTML]{FFFFFF} &
  \cellcolor[HTML]{FFFFFF} \\
\multirow{-2}{*}{VGG16} &
  \multirow{-2}{*}{\cellcolor[HTML]{FFFFFF}88.89\% \textcolor{Red}{(-2.18\%)}} &
  \multirow{-2}{*}{\cellcolor[HTML]{FFFFFF}0.7801} &
  \multirow{-2}{*}{\cellcolor[HTML]{FFFFFF}5.8450} &
  \multirow{-2}{*}{\cellcolor[HTML]{FFFFFF}0.0005} &
  \cellcolor[HTML]{C0C0C0}0.0833 &
  \cellcolor[HTML]{C0C0C0}15.5916 &
  \cellcolor[HTML]{C0C0C0}0.7181 &
  \multirow{-2}{*}{\cellcolor[HTML]{FFFFFF}0.0128} &
  \multirow{-2}{*}{\cellcolor[HTML]{FFFFFF}3,587} &
  \multirow{-2}{*}{\cellcolor[HTML]{FFFFFF}52,086} \\ \bottomrule
\end{tabular}}
\label{tab:var}
\end{table}

In contrast, SiftFunnel reduces MI by 50\% and lowers $\delta(z)$ to the millesimal scale, offering the strongest MIA resistance while maintaining usability. It achieves a minimum SSIM of 0.0793, highlighting its privacy–utility balance. Table~\ref{tab:var} compares performance across architectures.

\begin{figure}[t] 
    \centering
    \includegraphics[width=0.85\columnwidth]{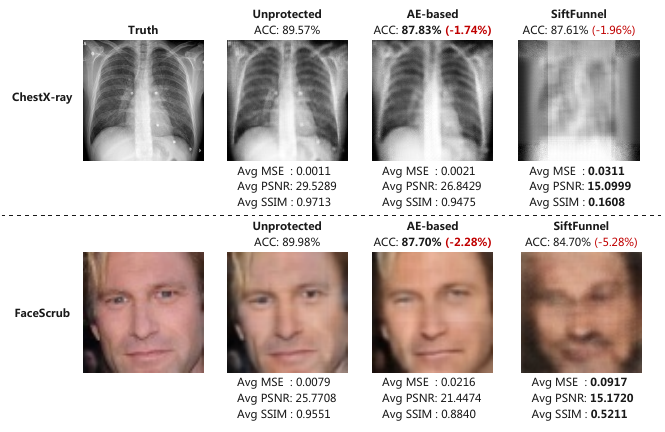}
    \caption{Quantitative and visual evaluation of privacy and usability against Gen-based MIA in the VGG16 edge model on the ChestX-ray ($\lambda_1=3.5,\lambda_2=1.2,\lambda_3=0.8$) and FaceScrub ($\lambda_1=4.5,\lambda_2=0.4,\lambda_3=0.1$).}
    \label{fig:cf}
    \vspace{-1.0em}
\end{figure}

\textit{3) The effect of data type. }Changes in data types lead to shifts in task requirements. For instance, in this study, the facial recognition task involves 530 classes, while ChestX-ray classification involves 4 classes, necessitating parameter adjustments as shown in Table~\ref{tab:face}. As analyzed in the previous section on model architectures, to achieve better defense performance, we adjusted the edge model divisions for ResNet-18 and VGG. Specifically, the division points were set after the second residual block for ResNet and after the first ReLU layer following the second max-pooling layer for VGG. The results, summarized in Table~\ref{tab:face} and Figure~\ref{fig:cf}, show that the defense performance remains unaffected by changes in data types. SiftFunnel continues to be the best defense method, offering an optimal balance between usability, privacy, and deployability.

As shown in Figure~\ref{fig:cf}, the VGG16 model with SiftFunnel shows better resistance to Gen-based MIA in terms of the visual effects of usability and privacy on facial and X-ray data, with PSNR values consistently around 15.

\textit{4) Ablation Study. }To assess the effectiveness of SiftFunnel’s model structure, loss function design, and defense capability, we conducted ablation experiments using Gen-based MIA. As shown in Table~\ref{tab:ablation}, we separately removed the Information Funnel and Attention Blocks from the model structure. The results indicate that removing the Funnel had minimal impact on accuracy, reducing it by only 0.1\%, whereas removing the attention mechanism significantly increased MI, weakening privacy protection. For the loss function, we examined the effect of removing KL divergence, linear and nonlinear constraints, and the $l_1$-norm. The results show that KL divergence and $l_1$-norm provided moderate defense improvements, while nonlinear constraints were critical—removing them increased MSE to 0.0268. Additionally, we validated the hypothesis from Section~\ref{sec:5.2}, demonstrating that even when adversaries reconstructed inputs using cloud-compensated features, SiftFunnel still outperformed existing defenses in protecting transmitted features.

We also evaluated the edge model’s latency on both CPU and GPU under the same configuration. For single-sample input, the unprotected, AE-based, and SiftFunnel-protected models had CPU latencies of 1.204 ms, 2.812 ms, and 2.192 ms, respectively. On GPU, their latencies were 3.9 ms, 4.3 ms, and 6.9 ms. Although SiftFunnel incurred slightly higher latency on GPU, it consumed significantly less memory, allowing more parallel input processing within the same resource constraints.



\section{Conclusion}
While CI enhances efficiency, it also exposes transmitted features to MIAs, posing severe privacy risks. Despite extensive research, a fundamental understanding of the factors determining MIA effectiveness remains lacking, and existing defenses struggle to balance usability, privacy, and deployability, facing theoretical, methodological, and practical constraints. To address these challenges, we proposed the first criterion for evaluating MIA difficulty in CI, identifying MI, entropy, and effective information volume as key influencing factors. Based on this, we proposed SiftFunnel, a privacy-preserving framework designed to limit redundant information while maintaining usability and deployability. By integrating linear and nonlinear correlation constraints, label smoothing, and a funnel-shaped edge model with attention mechanisms, SiftFunnel achieved robust privacy protection with an average accuracy loss of around 3\%, and revealed a nearly $20\times$ reduction in edge burdens. Experimental results demonstrated that SiftFunnel offers superior privacy protection across various evaluation metrics and achieves an optimal balance among usability, privacy, and practicality.
\bibliographystyle{IEEEtran}
\bibliography{hbip}
%



\end{document}